# Tapering to Enhance the Performance of Nanowire-Based Fluorescent Biosensors

Farhana Akter Mou* and Nicklas Anttu

*Physics, Faculty of Science and Engineering, Åbo Akademi University, FI-20500 Turku, Finland.*
**email: farhana.mou@abo.fi*

**Abstract:**

High-refractive-index nanowires are of interest for signal enhancement in fluorescence-based biosensing. Here, through electromagnetic modelling, we demonstrate substantial fluorescence signal enhancement by incorporating a tapered section into a cylindrical nanowire and investigate two complementary tapering configurations to address practical implementation challenges. Under near-normal excitation, the excitation enhancement increases approximately linearly with taper length, reaching more than three times that of the corresponding non-tapered nanowire at a taper length of 8000 nm. In a non-imaging signal detection scheme using a low numerical aperture (NA), the proposed tapered nanowire achieves more than three orders of magnitude enhancement in fluorescence signal compared with a planar biosensor. Thus, the tapered nanowire should enable efficient fluorescence detection using compact, low-NA optical configuration. In imaging mode, similarly as for non-tapered nanowires, imaging sharpness beyond conventional microscopy limitations is obtained when imaging from the non-tapered nanowire end, while strongly increased excitation enhancement is simultaneously obtained when exciting from the opposite, tapered end.

## 1. Introduction:

Biosensors have emerged as essential analytical tools owing to their unique ability to provide rapid, sensitive, and specific detection of target analytes.[1] Utilizing biological recognition elements to selectively interact with target molecules and transduce these interactions into quantifiable signals, biosensors offer high sensitivity and selectivity, thereby finding extensive applications in healthcare diagnostics, food safety assessment, and environmental monitoring[2]. Optical biosensors are of great interest because they offer exceptional sensitivity, allow label-free or fluorescence-based detection, real-time monitoring, and exhibit reduced susceptibility to interference from other biomolecules in the sample. These features make them well-suited for handling complex biological samples and performing multiplexed measurements[3,4]. Conversely, electrochemical, electrical, and piezoelectric sensors offer complementary advantages. For instance, electrochemical biosensors are typically inexpensive, working with straightforward instrumentation, and are highly suitable for miniaturization and portable devices[5]. Electrical biosensors can also achieve sensitive single-biomarker detection and offer the potential for compact and low-power sensing platforms[6]. Mechanical or piezoelectric sensors can detect mass changes directly, allowing label-free detection without the need for optical components[7]. However, compared to fluorescence-based optical biosensors, these platforms tend to be more susceptible to electrical noise, interference, and nonspecific interactions, offering less versatility in multiplexed assays, and generally face challenges in achieving ultra-high sensitivities[8].

Among the different optical biosensing platforms, planar biosensors have attracted considerable attention due to their simple architecture, ease of fabrication, and compatibility with surface functionalization strategies, making them highly suitable for integration into miniaturized and cost-effective sensing devices. Despite these advantages, planar biosensors suffer from inherent limitations, including a relatively low surface-to-volume ratio, weak signal-to-noise ratio in fluorescence-based detection, and limited sensitivity for low-abundance analytes, which hinder their broader applicability[9–11].

To address these challenges, nanowire biosensors are emerging as a promising solution, offering increased surface-to-volume ratio and efficient light–matter interaction, leading to superior sensitivity, thereby enabling the reliable detection of biomolecules at ultralow concentrations. Highly sensitive fluorescence-based nanowire biosensors could find use for example for medical diagnosis of early-stages of cancer[12], asthma[13], and heart disease[14]. Semiconducting nanowires, commonly composed of materials such as GaP[15], GaAs[16], ZnO[17], Si[18], and InAs[19], exhibit unique optical properties, including high refractive index and strong light confinement, making them promising platforms for fluorescence-based biosensing. Numerous investigations have explored the capability of nanowires to enhance fluorescence signals, with several specifically investigating interactions between nanowires and fluorescent dyes used for biomolecular labeling[15–18]. For example, in Ref.[20], $SiO_2$-coated GaP nanowires were used to detect protein biomarkers in highly diluted human serum. Compared to conventional flat silica or plastic substrates, the nanowire platform enhanced signal intensity, contrast, and signal-to-noise by over tenfold, demonstrating improved sensitivity and lower detection limits for standard diagnostic assays.

The research reported in Ref.[21] demonstrated how the efficiency of light guiding in nanowires is influenced by the choice of material, diameter, and light wavelength for excitation ($\lambda_{exc}$) and emission ($\lambda_{em}$), where certain diameters achieve optimal efficiency across specific wavelength ranges, facilitating tailored optimization for specific fluorophores. Furthermore, the placement of the nanowire (horizontal vs. vertical) can have a strong effect on the optical response. Specifically, arrays of vertical nanowires offer large surface area and ordered layout that increase sensitivity and consistency. Meanwhile, the extensive array coverage enables easy wide-field imaging for quick and multiplexed analysis with standard fluorescence microscopy[22]. Considering these factors, several nanowire biosensors have been proposed, and their performances have been systematically analyzed by different researchers[15–24].

Recently, our research group showed through modelling how vertically placed GaP cylindrical shaped nanowires can enhance fluorescence excitation intensity, improve photon collection efficiency, and increase quantum yield through the Purcell effect, leading to over 100× signal enhancement compared to conventional planar biosensors[23] . Importantly, high performance at low numerical aperture (NA: $NA_{exc}$ and $NA_{coll}$ for excitation and fluorescence collection, respectively) operation was shown, which is particularly advantageous, as it opens the door to low-cost, compact optical components. Additionally, the work highlights the advantage of nanowires in imaging-based detection, allowing for 25 times sharper images, beyond conventional microscopy limitations, making them promising for portable, high-sensitivity biosensing applications[23].

Here, we develop and apply modelling approaches for the design and analysis of tapered nanowire-based fluorescent biosensors, focusing on the five optical effects illustrated in Figure 1.

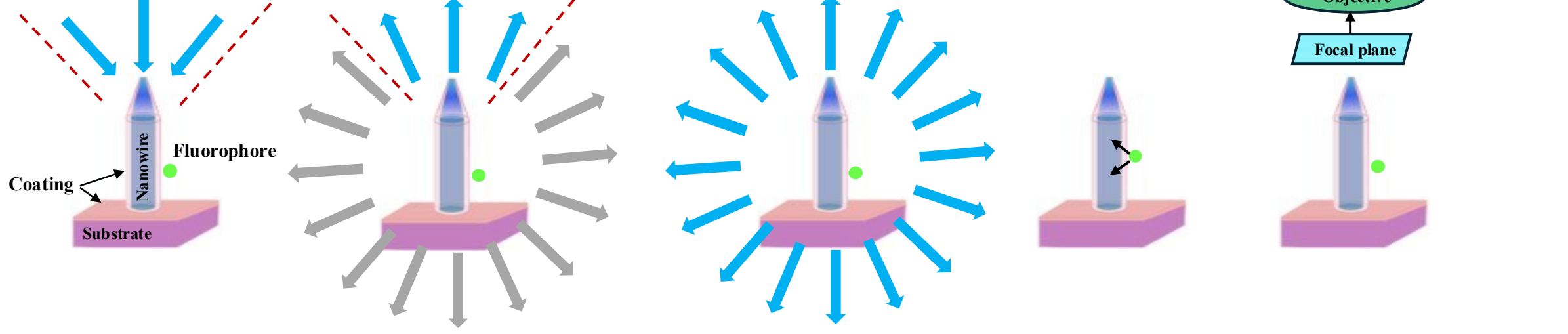

Figure 1: Schematic of (a-d) the four optical effects and (e) fluorescence imaging-detection concept studied here.

In our design, a tapered section is integrated with the cylindrical nanowire to enhance light concentration, while the fixed cylindrical base allows to retain the previously found[23] superior imaging performance. The optimized tapered nanowire design can direct excitation light to the targeted area, while also enhancing collection efficiency. This interaction thus further amplifies the fluorescence signal and enhances the performance of the biosensor.

For biosensors working in imaging mode, we show that by placing the tapered section into the substrate, with bottom-side excitation through the substrate and top-side imaging, strong tapering-induced excitation enhancement is obtained, while imaging clarity is retained from the non-tapered cylindrical part.

At low numerical aperture for excitation and collection ($NA_{exc}$ and $NA_{coll}$) the proposed tapered nanowire can exhibit more than an order of magnitude improvement in signal compared to the non-tapered counterpart, and the improvement increases with increasing tapering length. Thus, a signal enhancement by more than three orders of magnitude, compared to the planar biosensor, is possible. This behavior highlights the potential of tapered nanowires as a cost-effective biosensor that can be easily integrated into compact optical systems.

## 2. Materials and methods:

Here, we focus on Si as a material for the nanowire. Previously, it was shown that Si and for example GaP perform well for nanowire-based biodetectors[23]. Si is more cost-effective than GaP, but shows some parasitic absorption, especially at short wavelengths in the visible wavelength range. At longer wavelengths, when not too limited by parasitic absorption, Si can outperform GaP due to the higher (real part of the) refractive index of Si[23]. For completeness, we also investigate the signal enhancement performance of tapered GaP nanowires in Section 2 of the Supporting Information, with similar overall conclusions as for the Si nanowires below.

The optical response of the tapered nanowires was modelled with the finite element method in COMSOL Multiphysics, see Supporting Information Section 1 and Ref.[23] for technical details.

The proposed nanowire features a dual-sided architecture, integrating a cylindrical section that preserves the imaging performance of the non-tapered nanowire[23] with a tapered section engineered to further enhance the signal, as illustrated in Figure 2. We investigate two nanowire configurations, namely top-tapered (Figure 2 (a)) consisting of a vertically oriented Si nanowire integrated on a Si substrate and bottom-tapered (Figure 2(b)) where the tapered section is positioned within a lower refractive index substrate, such as a plastic substrate with $n \approx 1.5$), considering several practical and performance-related factors (see Sections 2.2 and 2.3 for discussion). These factors include minimizing fabrication complexity, eliminating the need for selective functionalization, facilitating operation in the imaging mode, and achieving a favorable balance between signal enhancement and imaging performance. The rationale for selecting the optimal configuration and its corresponding performance characteristics are discussed in detail in the following sections.

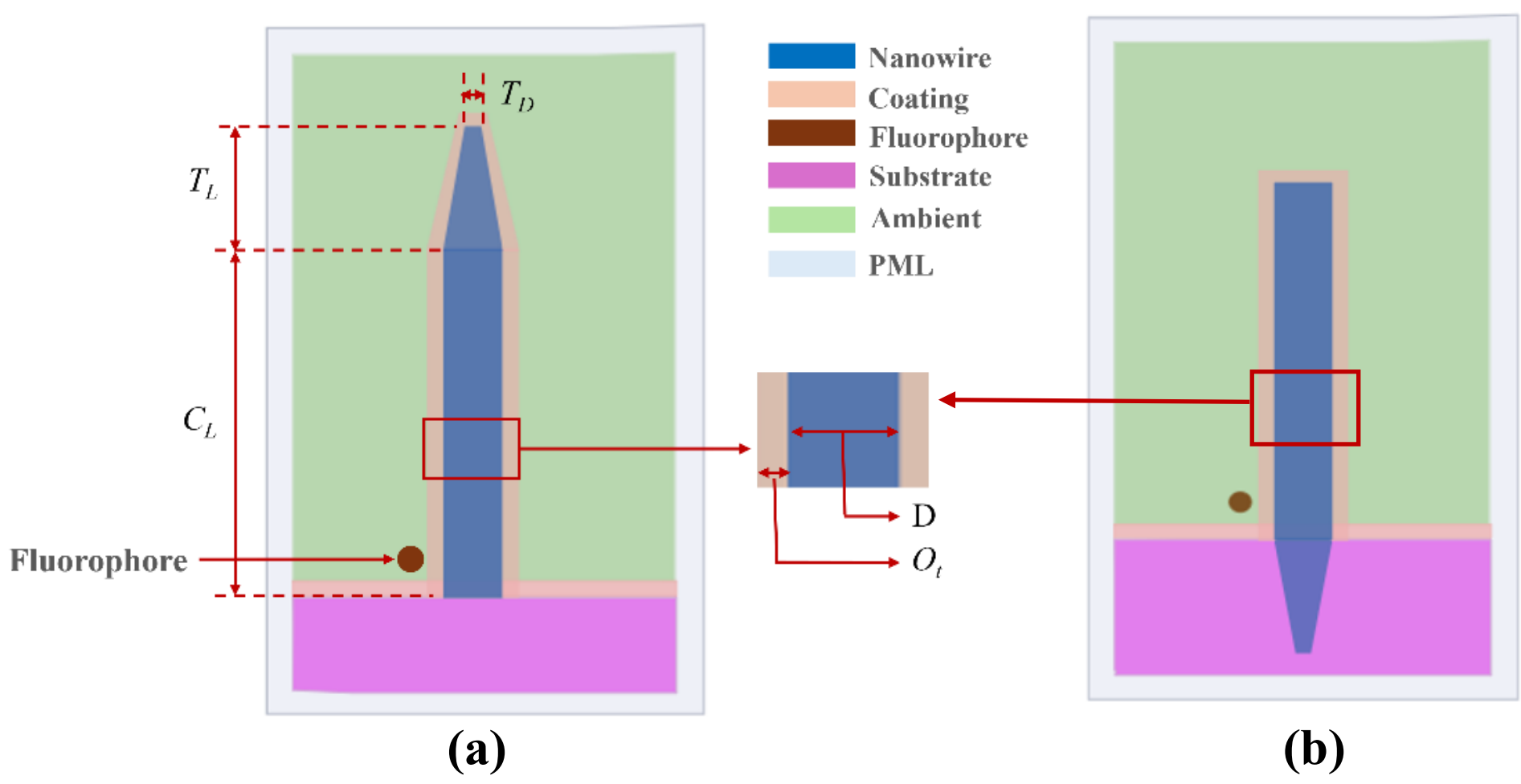

Figure 2: Proposed (a) top-tapered and (b) bottom-tapered nanowire configurations together with materials and geometrical parameters indicated.

The length $C_L$ for the cylindrical section was chosen to 1500 nm, long enough that the performance of the non-tapered nanowires were found to become practically independent of the nanowire length[23]. The nanowire diameter $D$ is considered in the range 50 to 250 nm. The length of the tapered section ($T_L$) is varied from 500 nm to 8500 nm, while the tip diameter is fixed at a small value of $T_D$ = 20 nm. An $O_t$ = 10 nm thick $SiO_2$ layer, providing a chemically functionalizable surface for selective biomolecule attachment, is on top of the nanowire. A fluorophore is placed at a distance of 5 nm from the nanowire sidewall, where its axial binding location can vary[23]. We considered the wavelengths spanning from 400 to 900 nm, a broad spectral window that encompasses the emission and excitation bands of many fluorophores commonly used in practical applications[23]. We performed our simulation within a homogeneous surrounding medium of water ($n \approx 1.33$), a pragmatic choice to simplify calculations while accurately reflecting biologically relevant conditions encountered in biosensing and bioimaging applications. For the $SiO_2$, we use tabulated values[25] of $n \approx 1.45$, and for Si tabulated values from Ref.[26] (with Re($n$) ≈ 3.5 - 5.5, depending on wavelength). To eliminate spurious reflections from the external boundaries of the simulation domain, we included perfectly matched layers (PMLs) that absorb outgoing electromagnetic waves without spurious back-reflection[23].

### 2.1: Signal Enhancement Model:

Fluorescence-signal enhancement in nanowire systems is shaped by a complex interplay of optical effects arising from the local electromagnetic environment[23], which we summarize below for self-consistency (see Figure 1 for schematic). Specifically, the emission signal of a fluorophore is influenced by the local enhancement of the excitation field, $ENH_{exc}$, the enhancement of the efficiency with which emitted photons are collected, $ENH_{coll}$, modifications to the intrinsic emission rate, quantified by the Purcell factor, $C_{Purcell}$, and possible parasitic absorption losses, $A_{parasitic}$. Each of these contributions depends sensitively on the nanowire's material composition, geometry, and the precise location of the fluorophore along the nanowire. All enhancement factors are relative to a fluorophore in a homogeneous medium with refractive index $n$ = 1.33, that is, close to that of $H_2O$ and identical to the surrounding environment in our nanowire modelling. Additional technical details are given in Section 1 of the Supporting Information**.**

$ENH_{exc}$ is affected by the wavelength of the incident light and the illumination configuration. For wide-field excitation, the local excitation intensity, and hence $ENH_{exc}$, is obtained by incoherently integrating over all angles within the excitation numerical aperture $NA_{exc}$[23]. Likewise, $ENH_{coll}$ depends on the emission wavelength and is evaluated by integrating over the emission angles accepted by the collection optics, defined by $NA_{coll}$, under the assumption that every photon entering this angular range contributes to the detected signal (in Section 3.2, we discuss the complementary imaging-based detection). In contrast, both the Purcell factor and parasitic absorption are independent of the illumination and collection optics, but depend on emission wavelength.

The Purcell effect modifies the fluorophore's intrinsic quantum yield ($QY_0$), giving rise to a position-dependent quantum yield modification[23]:

$$\mathrm{ENH_{QY}} = \frac{C_{\mathrm{Purcell}}}{C_{\mathrm{Purcell}}\,\mathrm{QY_0}+(1-\mathrm{QY_0})}. \quad (1)$$

In the low-excitation regime, where excitation-saturation effects are negligible, the overall fluorescence signal scales linearly with the product of these factors[23]:

$$\mathrm{ENH_{sig}} = ENH_{\mathrm{exc}} \times \mathrm{ENH_{coll}} \times \mathrm{ENH_{QY}}. \quad (2)$$

A signal enhancement of unity, that is, $\mathrm{ENH_{sig}} = 1$, corresponds to the signal from a fluorophore in the homogeneous reference medium, when probed with the same $\mathrm{NA_{exc}}$ and $\mathrm{NA_{coll}}$ as for the nanowire biosensor. Parasitic absorption reduces the collection efficiency, thereby diminishing the overall signal enhancement[23].

Since fluorophores can occupy different positions along the nanowire, we define an axially average signal enhancement:

$$\langle \mathrm{ENH_{sig}} \rangle = \langle \mathrm{ENH_{exc}} \times \mathrm{ENH_{coll}} \times \mathrm{ENH_{QY}} \rangle \quad (3)$$

where the angled brackets denote averaging over the fluorophore's axial position. This approach provides a unified framework to describe how local excitation, photon collection, Purcell factor, and absorptive losses collectively shape the observed fluorescence in nanowire systems. Similarly, with $\langle \mathrm{ENH_{exc}} \rangle$, $\langle \mathrm{ENH_{coll}} \rangle$, and $\langle \mathrm{ENH_{QY}} \rangle$, we define the respective axially averaged enhancement factors.

### 2.2: Selective functionalization for top-tapered configuration:

We find that excitation enhancement for fluorophores that attach to the side wall of the tapered section tends to be much weaker than for fluorophores that bind to the side wall of the cylindrical section (see Figure 3(a) for $D$ = 100 nm, which is a rather optimized diameter, as seen from the diameter dependence in Figure 3(b)). Thus, to maximize the signal, selective functionalization is required to ensure that fluorophores bind only to the cylindrical region. Therefore, below, when calculating all the enhancement factors for the top-tapered configuration, we considered binding of fluorophores only to the cylindrical section of $C_{\mathrm{L}}$ = 1500 nm.

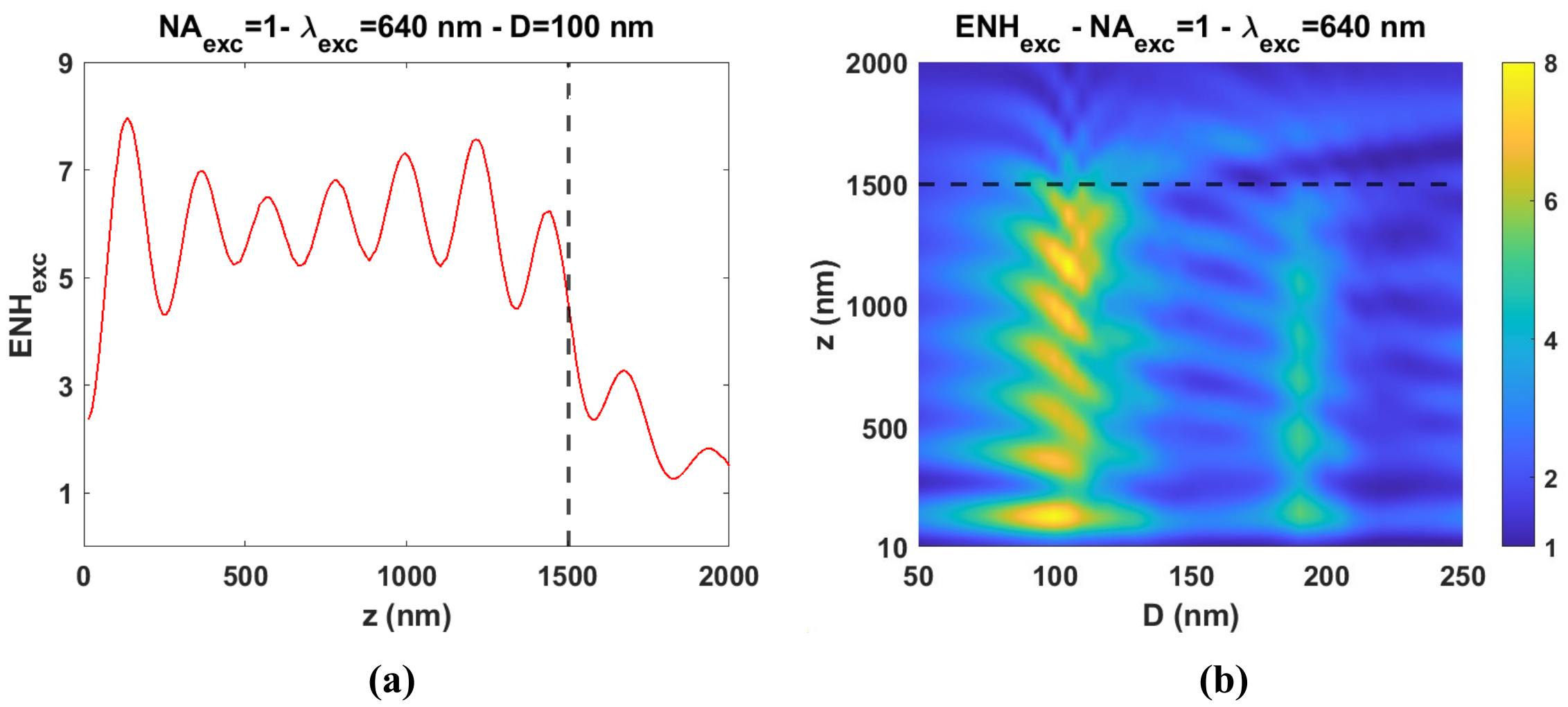

Figure 3: Enhancement of the incident light intensity for fluorophores located along the sidewalls of the top-tapered design (see Figure 2(a) for schematic), (a) as a function of axial position for $D$ = 100 nm, with $z$ = 0 corresponding to the substrate surface and $z$ = 2000 nm to the top of the nanowire, where the top-most 500 nm is the tapered section (indicated by the dashed line) and (b) same as (a) but also with diameter dependence. Here, $\mathrm{NA_{exc}}$ = 1 and $\lambda_{\mathrm{exc}}$ = 640 nm.

**2.3: Inverted bottom-tapered configuration:**

If achieving the above selective functionalization turns out technically demanding, we propose as an alternative an inverted configuration (see Figure 2 (b)) with tapering of the nanowire into the substrate (which, for the light guiding to work in the nanowire, then needs to be of lower refractive index than the nanowire material, for example a plastic substrate of $n \approx 1.5$). Then, the substrate will prevent fluorophores from binding to the low-enhancement tapered region, without the need for selective surface functionalization. Such fabrication of tapered nanowires into the substrate is technically feasible, as controlled tapered nanostructures can be fabricated using established methods such as multi-step etching[27]. A discussion of the feasibility of fabricating the bottom-tapered configuration is found in Section 1-B in the Supporting Information. Below, we focus on the top-tapered configuration and include main results also for the bottom-tapered nanowire for comparison [See Section 1-B in the Supporting Information for additional results for the inverted (bottom-tapered) configuration].

**3. Results: Numerical Optical Responses and Analysis**

We focus on $\lambda_{exc}$ = 640 nm and $\lambda_{em}$ = 670 nm, which is a typical configuration for experiments with Alexa Fluor 647 fluorophore[28]. To generalize beyond that fluorophore system, we show details of the wavelength-dependence in the Supporting Information.

**3.1(a): Excitation Enhancement:**

Figure 4(a) illustrates the axial dependence of excitation enhancement $ENH_{exc}$ for Si top tapered nanowires with diameter $D$ = 100 nm [which is a rather optimized diameter as seen in Figure 4(b)], excitation wavelength $\lambda_{exc}$ = 640 nm, and $NA_{exc}$ = 0.1 for tapering lengths $0 \leq T_L \leq 8500$ nm. We find an increasing $ENH_{exc}$ with increasing tapering length, which we assign to tapering-assisted enhanced excitation of the $HE_{11}$ mode for the close to normal incidence excitation at this low $NA_{exc}$ [23], as discussed in more detail below. The peak enhancement systematically rises with tapering length, from about 30 without tapering to around 75 for $T_L$ = 8500 nm. An oscillatory axial pattern, set by interference of incident and silicon-substrate-reflected light remains unchanged by $T_L$, with dip values on the order of 50% of the peak values. We note that the inverted bottom-tapered geometry has as a possible benefit a reduced visibility of such oscillatory pattern (Figure S14 in the Supporting Information) thanks to the lower refractive index substrate in that case.

Tapered nanowires exhibit a nearly linear improvement in excitation enhancement with increasing tapering length at low $NA_{exc}$, for both top and bottom-tapered configurations, as presented in Figure 4(d). This trend we explain as follows. The small-diameter tip of the tapering allows to efficiently capture incident light from a large surrounding volume, with adiabatic funneling of the light into the guided $HE_{11}$ mode of the nanowire. Increasing the tapering length improves this adiabatic mode matching between the incident free-space plane wave and the guided mode in the cylindrical part of the nanowire, thereby enhancing coupling efficiency and producing a stronger localized excitation field at the nanowire side wall where the fluorophores reside. This interpretation is further supported by the electric field intensity distributions shown in Figure S3 in the Supporting Information for taper lengths of 500, 2000, 4000, and 8000 nm. There, with increasing taper length, the optical field becomes progressively more concentrated near the nanowire sidewall, while a low-intensity shadow region emerges in the surrounding medium. This shadow indicates that a larger fraction of the incident optical power from the surrounding of the nanowire is transferred into the nanowire-guided mode, providing direct visual evidence of the improved adiabatic coupling into the guided mode with increasing $T_L$. This adiabatic electromagnetic funneling to the $HE_{11}$ mode with a very small diameter tip at the top of the tapering thus uses the opposite geometry compared to the expectation of a light funnel in ray-optics (see Figure S4 in the Supporting Information for schematic).

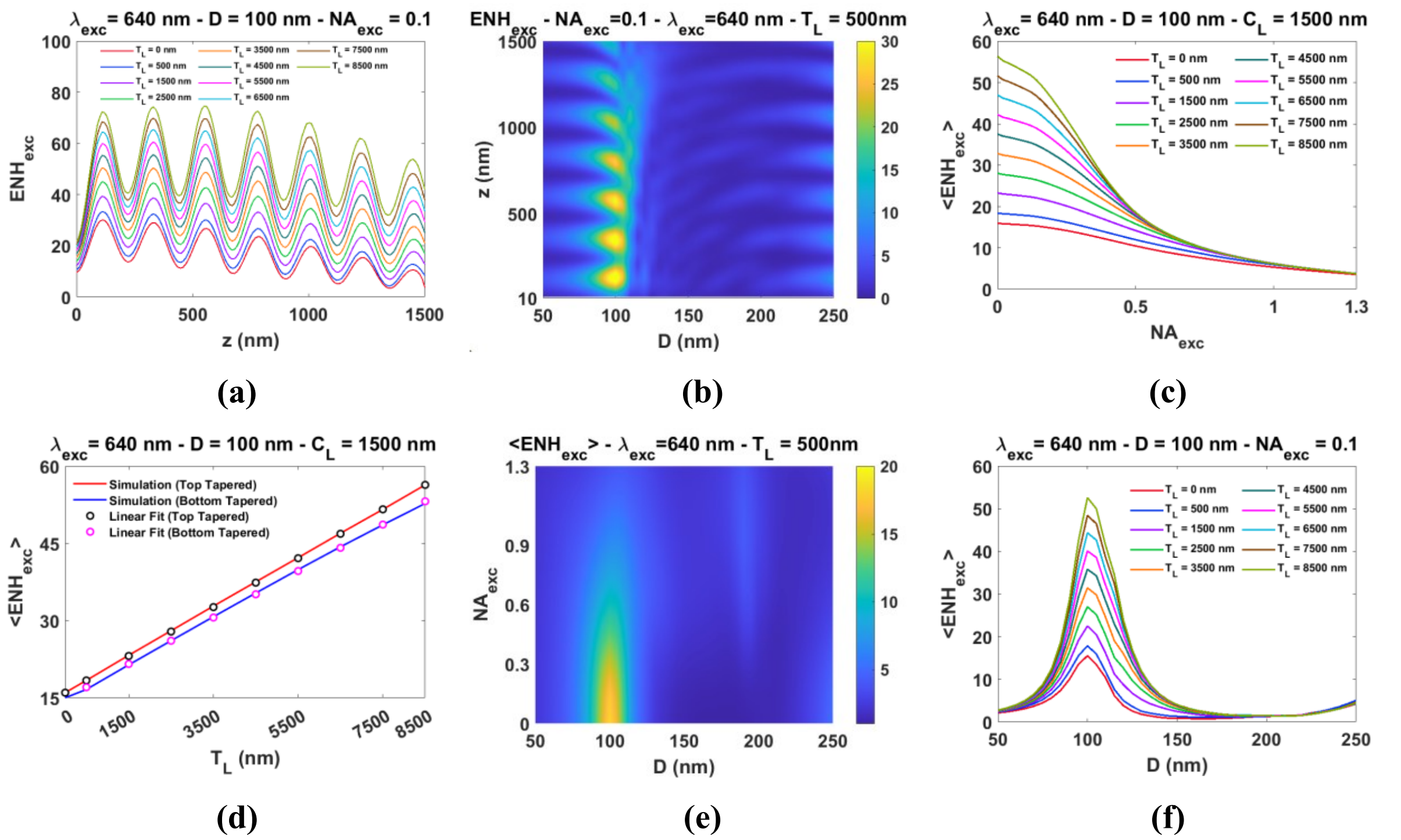

Figure 4: Enhancement of the incident light intensity for the top-tapered configuration (except for (d) that shows for both configurations—overall, the bottom-tapered configuration exhibits a similar excitation-enhancement dependence to this top-tapered configuration (see the Supporting Information**,** Section 1 and Figure S14). (a) The variation of $ENH_{exc}$ at $\lambda_{exc} = 640$ nm, $NA_{exc} = 0.1$, and $D = 100$ nm as a function of axial position of top-tapered nanowire, with $z = 0$ corresponds to the substrate surface and $z = 1500$ nm denotes the top of the cylindrical binding region (i.e., the onset of the tapered section). (b) Same as (a) but also with diameter dependence. (c) $\langle ENH_{exc}\rangle$ of top-tapered configuration as a function of $NA_{exc}$, for different tapering lengths. (d) $\langle ENH_{exc}\rangle$ as a function of tapering length for $NA_{exc} \rightarrow 0$ for both top- and bottom-tapered configuration (with excitation incident toward the tapered side in both configurations) together with linear fit to the simulated data. (e) same as (c) but also with diameter dependence. (f) $\langle ENH_{exc}\rangle$ as a function of $D$ and $T_L$.

From Figure 4(d), we find from linear fitting that the excitation enhancement varies as $A + B \times T_L$ with $A = 16$ and $B = 0.0047$ nm$^{-1}$ for the top tapered configuration and $A = 15$ and $B = 0.0044$ nm$^{-1}$ for the bottom tapered configuration. The linear increase in the large range of $T_L$ studied here suggests that further increases in $T_L$ should yield additional linear enhancement.

Figure 4(f) shows that the excitation enhancement decreases only slightly if the diameter varies by ±10 nm around the optimal nanowire diameter, thus giving room for minor variations in the fabrication of the nanowires. For all the tested tapering length cases, the same behavior is observed, and the peak value is enhanced with the tapering length without shifting the optimal diameter. The optimum diameter depends on $\lambda_{exc}$ and optimal diameter varies almost linearly with the excitation wavelength $\lambda_{exc}$ (see Figure S5 in the Supporting Information). Such a linear behavior is expected if the refractive indices of the constituent materials are assumed to be wavelength independent[28].

Figure 4(c) highlights the strong dependence on $NA_{exc}$ for varying $T_L$. The highest enhancement occurs at low $NA_{exc}$ where excitation is concentrated near normal incidence, at which the most efficient coupling into the $HE_{11}$ mode is expected[23]. Here, $\langle ENH_{exc}\rangle$ at low $NA_{exc}$ reaches values of approximately 56 for $T_L$ = 8500 nm compared to about 16 for the non-tapered ($T_L$ = 0 nm) nanowire, in good agreement with the linear fitting $A + B \times T_L$ from above. Figure 4(e) shows that the maximum enhancement is concentrated around $\theta = 0°$ (that is, when $NA_{exc} \rightarrow 0$ where the $HE_{11}$ mode is efficiently excited) for a nanowire diameter of $D$ = 100 nm. At this $D$, the $HE_{11}$ waveguide mode shows in its modal pattern a large field strength at the (side) surface of the nanowire where the fluorophores reside, thus giving the diameter-dependent peak in $\langle ENH_{exc} \rangle$. Thus, lowering the $NA_{exc}$ markedly increases the average excitation enhancement $\langle ENH_{exc} \rangle$, from approximately 6 at $NA_{exc}$ = 1 to about 19 at a low $NA_{exc}$ for $T_L$ = 500 nm, as observed in both Figure 4 (c, e), similarly as previously reported for non-tapered nanowires[23].

**3.1(b): Collection Enhancement:**

Also fluorophore emission collection shows a strong dependence on NA. The maximum collection enhancement is noticed at around $\theta = 0°$ (Figure 5(a)), again for a nanowire diameter of $D$ = 100 nm (see Figure S6 and Figure S15(b) in the Supporting Information). For instance, at $NA_{coll}$ = 1, $\langle ENH_{coll} \rangle$ reaches approximately 1.95, while at $NA_{coll} \rightarrow 0$, it rises to around 5.5 for a tapering length $T_L$ = 500 nm. It is also noticeable that at such a low $NA_{coll}$, increasing tapering length enhances the $<EHN_{coll}>$ significantly. Similarly, as for the excitation enhancement above, we find that linear fitting to the data in Figure 5(b) works well, yielding $\langle ENH_{coll} \rangle \approx C + D \times T_L$ with $C$ = 4.4 and $D$ = 0.0012 $nm^{-1}$ for top-tapered configuration and $C$ = 4.4 and $D$ = 0.0013 $nm^{-1}$ for bottom-tapered configuration. Thus, for example, when adding an 8500 nm tapering on top of a 1500 nm cylindrical nanowire, the $<ENH_{coll}>$ increases by more than 3 times compared to the non-tapered nanowire ($T_L$ = 0 nm).

Such a similar linearly dependent type of increase here in $\langle ENH_{coll} \rangle$ as in $\langle ENH_{exc} \rangle$ in Figure 4(d) could actually be expected from reciprocity between in-coupling and out-coupling of light, as given by the Lorentz reciprocity[29]. Reciprocal to the light funneling of excitation light, we assign this behavior in emission as follows. A longer taper functions as a light-directing antenna, coupling those photons emitted into the $HE_{11}$ mode to a narrow exit angle (which decreases with increasing tapering length, where in the limit of infinitely long tapering, the $HE_{11}$ mode transforms adiabatically into a plane wave propagating parallel to the direction of the nanowire axis, corresponding to emission angle of $\theta = 0$), thus enhancing photon collection at low $NA_{coll}$.

However, increasing $\langle ENH_{coll} \rangle$ by reducing $NA_{coll}$ leads to a decrease in the absolute signal while simultaneously enhancing the signal relative to that from a reference fluorophore in a homogeneous liquid collected under the same $NA_{coll}$. In other words, while reducing $NA_{coll}$

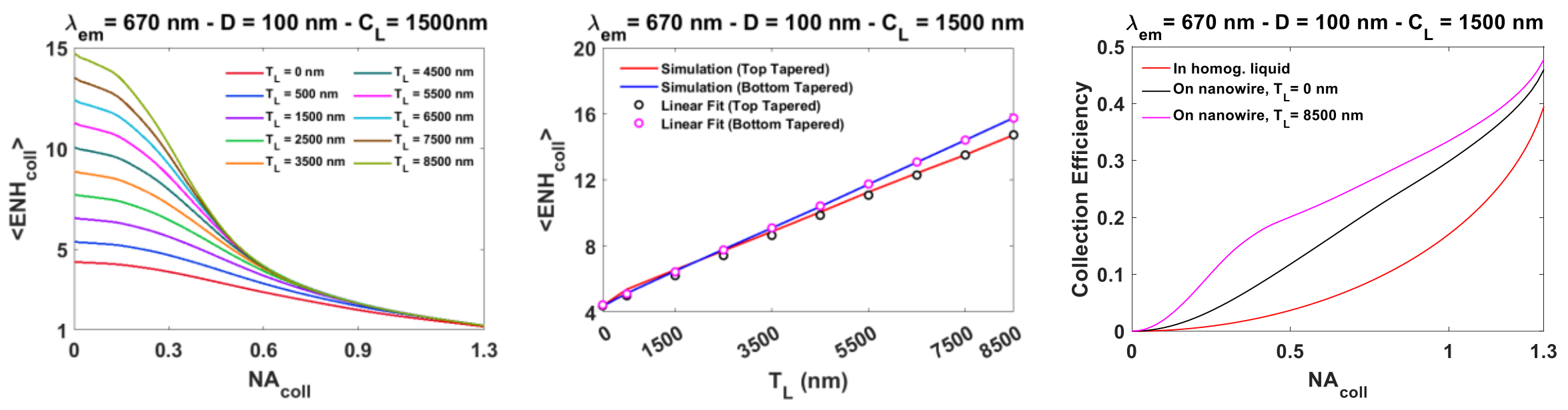

**(a)** **(b)** **(c)**

Figure 5: Enhancement of fluorophore emission collection for $C_L$ = 1500 nm, $D$ = 100 nm, $O_t$ = 10 nm, and $\lambda_{em}$ = 670 nm. (a) $\langle ENH_{coll} \rangle$ as a function of $NA_{coll}$ for top-tapered configuration. (b) $\langle ENH_{coll} \rangle$ as a function of tapering length for $NA_{coll} \rightarrow 0$ for both top- and bottom-tapered configuration (with collection from the tapered side in both configurations) together with linear fitting to the simulated data. (c) Collection efficiency as a function of $NA_{coll}$, for the top-tapered nanowire and the homogeneous test liquid.

increases ⟨$ENH_{coll}$⟩, it lowers the absolute signal. Importantly, the increase in ⟨$ENH_{coll}$⟩ can improve the signal-to-noise ratio if the dominant noise is optical in origin and uniformly distributed across all angles within the collection objective.

For instance, in the case of top tapered configuration, in a homogeneous liquid, 17.9% of photons are collected at $NA_{coll}$ = 1.0, compared to just 0.14% at $NA_{coll}$ = 0.1. On the other hand, with a non-tapered nanowire ($T_L$ = 0 nm), about 29.9% of emitted photons are collected at $NA_{coll}$ = 1.0, but only 0.61% at $NA_{coll}$ = 0.1, and in the case of tapered nanowire with $T_L$ = 8500 nm, 33.5% of emitted photons are collected at $NA_{coll}$ = 1.0 and 1.98% at $NA_{coll}$ = 0.1. These situations are displayed in Figure 5 (c).

We can summarize this behavior by noting that at high $NA_{coll}$ of 1.0, both non-tapered and tapered nanowires provide an approximately two-fold improvement in collection efficiency compared to a homogeneous liquid. However, at low $NA_{coll}$, this improvement becomes much more pronounced: nearly five-fold for the non-tapered nanowire and fourteen-fold for the tapered nanowire (for $T_L$ = 8500 nm). Importantly, the results in Figure 5(c) show that equivalent photon collection can be achieved at low NA with a tapered nanowire compared to a conventional planar biosensor operating at high NA. For instance, the 17.9% photon collection obtained at $NA_{coll}$ = 1.0 in a homogeneous liquid can be achieved at $NA_{coll}$ = 0.4 using a tapered nanowire with $T_L$ = 8500 nm. This increase highlights the potential of the proposed tapered nanowire architecture to achieve high photon collection efficiency without the need of high NA read-out setup.

### 3.1 (c): Purcell factor and parasitic absorption

The Purcell factor for the top-tapered configuration as function of diameter for $\lambda_{em}$ = 670 nm is shown in Figure 6(a). It can be observed that a Purcell factor greater than unity is achieved for all diameters and tapering configurations. The peak Purcell factor occurs at a diameter of $D$ = 120 nm, which is slightly shifted compared to the optimized diameter of $D$ = 100 nm for ⟨$ENH_{coll}$⟩ at the same wavelength, a general characteristics of the system, as seen from Figure S8 and Figure S16 (b) that show dependence also on the wavelength for top- and bottom-tapered configuration, respectively. Neither the peak position nor the peak values are shifted noticeably with increasing tapering length [Figure 6(a)]. A very similar diameter dependence for the Purcell factor is observed for the bottom-tapered design (Figure S16(a) in the Supporting Information).

We have evaluated the Purcell factor for $0 \leq T_L \leq 8500$ nm at optimum diameter for both top- and bottom-tapered configuration [Figure 6 (b)]. The Purcell factor is almost independent of $T_L$. This result shows that the Purcell factor is mainly set by the coupling of the fluorophore emission to the optical modes of the nanowire, possibly with contribution from back-reflection from the substrate surface, but without noticeable effects from reflection back from the tapered end of the nanowire, even for the non-tapered nanowire (and any such reflection is diminished with increasing tapering length). The important implication is then that $ENH_{QY}$ does not show noticeable dependence on the tapering length.

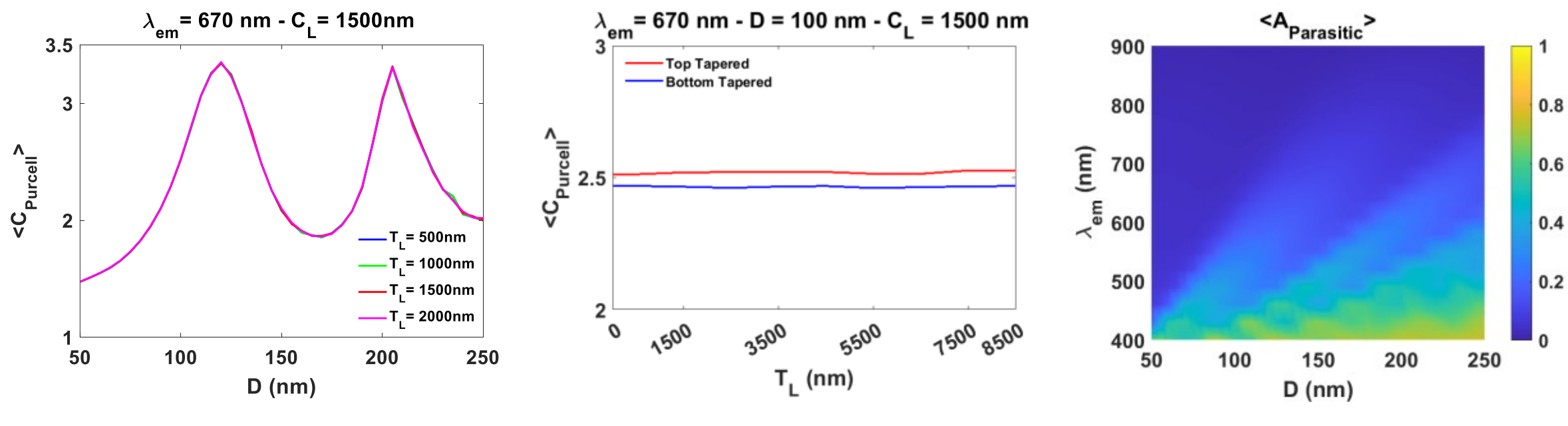

(a) (b) (c)

Figure 6: (a) Purcell factor for different tapering length as a function of diameter for top-tapered configuration. (b) Purcell factor of tapered nanowire with respect to tapering length for both top- and bottom-tapered configuration. (c) Parasitic absorption of emission from the fluorophores at the sidewalls of top-tapered nanowires for $T_L$ = 500 nm, $D$ = 100 nm, and $O_t$ = 10 nm as a function of emission wavelength and nanowire diameter.

Parasitic absorption is seen quite strongly at short wavelengths [Figure 6(c) for top-tapered configuration and Figure S16(c) in the Supporting Information for bottom-tapered configuration]. For example, in the top-tapered configuration at $\lambda_{em}$ = 500 nm, where the absorption length in bulk Si is less than 900 nm, around 30% of the emitted photons are lost because of parasitic absorption. On the other hand, at the wavelength $\lambda_{em}$ = 670 nm that we focused on above, the absorption length of bulk Si is approximately 4200 nm. There, less than 5% of the emitted photons are lost due to parasitic absorption for all our tested tapering lengths at the optimized diameter of 100 nm (see Figure S9 in the Supporting Information). A very similar absorption behaviour is also observed for the bottom-tapered nanowire, as shown in Figure S16(d) in the Supporting Information.

### 3.1 (d): Signal Enhancement:

High quantum yield fluorophores are preferred because they produce bright and strong signals. However, lower quantum yield fluorophores are sometimes necessary when specific excitation or emission wavelengths are required. Such fluorophores may also be selected if they show a greater resistance to photobleaching, which ensures more stable signal during long experiments. Therefore, it is crucial to understand how the fluorescence signal varies with fluorophores of different quantum yield, as this knowledge allows for proper selection of a dye that provides sufficient signal strength for reliable detection. Considering these necessities, in our case we have evaluated the signal enhancement with three different situations of quantum yield (here defined relative to the test liquid, before the modification due to the Purcell factor): a very low value of $QY_0 \rightarrow 0$, an intermediate value of $QY_0 = 0.33$, and a high value of $QY_0 = 1$.

Since we showed above that $C_{\text{Purcell}}$, and hence $ENH_{QY}$, is almost completely independent of tapering length, $ENH_{sig}$ varies as the product of $ENH_{exc}$ and $ENH_{coll}$. We note that for the above $\lambda_{exc}$ = 640 nm and $\lambda_{em}$ = 670 nm, thanks to the broad peak in both $ENH_{exc}$ and $ENH_{coll}$ as function of diameter discussed above, we can obtain a nearly optimum $EHN_{exc}$ and $EHN_{coll}$ with a given, optimized $D$.

From Figure 7(a), for top-tapered geometry at the optimized $D$ = 100 nm, $C_L$ = 1500 nm, and $T_L$ = 500 nm, for low NA ($NA_{exc} = NA_{coll} = 0.1$), we find for $QY_0 \rightarrow 0$ that $\langle EHN_{sig} \rangle = 265$, for $QY_0 = 0.33$ that $\langle EHN_{sig} \rangle = 174$, and for $QY_0 = 1$ that $\langle EHN_{sig} \rangle = 103$. We note that the ratio 265/103, which corresponds to the ratio of QY enhancement for $QY_0 \rightarrow 0$, equivalent to $C_{\text{Purcell}}$, to QY enhancement for $QY_0 = 1$, equivalent to 1, indeed agrees well with the $C_{\text{Purcell}} = 2.5$ at $D$ = 100 nm seen in Figure 6(a).

Importantly, the signal enhancement consistently increases with tapering length for all quantum yield cases, as shown in Figure 7(b) for low values of $NA_{exc}$ and $NA_{coll}$ for both top- and bottom-tapering configuration. For instance, for the top-tapered configuration with $QY_0 \rightarrow 0$, the signal enhancement increases from the above 265 at tapering length of 500 nm to 1909 when the tapering length is extended to 8500 nm.

For the system in Figure 7 (b), the $\langle ENH_{QY} \rangle$ at $QY_0 \rightarrow 0$, 0.33, and 1 is 2.52, 1.71, and 1 at $T_L$ = 0 (and we know that it is rather independent of $T_L$). Then, the $\langle ENHsig \rangle$ is given to a good approximation by $(A + B \times T_L) \times (C + D \times T_L) \times E/F/G$ respectively, where $E = \langle ENH_{QY} \rangle$ at $QY_0 \rightarrow 0$, $F = <ENH_{QY}>$ at $QY_0 = 0.33$, and $G = <ENH_{QY}>$ at $QY_0 = 1$—the agreement is limited then by how well $<ENH_{exc}ENH_{coll}ENH_{QY}>$ can be approximated by $<ENH_{exc}><ENH_{coll}><ENH_{QY}>$, which was shown to work well for $T_L = 0$ previously[23]. With the tapering, we expect the agreement to be even better since the tapering decreases the interference fringe-effects in the axial direction. Indeed, as seen in Figure S18 in the Supporting Information, the agreement between $(A + B \times T_L) \times (C + D \times T_L) \times E/F/G$ and the simulated $<ENH_{sig}>$ is good. Thus, now we can explain why we see for $<EHN_{sig}>$ a $T_L^2$ dependence in Figure 7(b): At low NA for excitation and collection, both $<ENH_{exc}>$ and $<ENH_{coll}>$ show a $T_L$ dependence, resulting in a $T_L^2$ dependence due to their product in $<ENH_{sig}>$.

To compare the performance to a conventional planar biosensor, we use values from a previous study[23] for a fluorophore located on a planar substrate coated with a 10 nm thick $SiO_2$ layer, identical to that used for the nanowire here. For the planar reference, we find an average signal enhancement factor $\langle ENH_{sig} \rangle$ in the range of 0.7-0.8 for all values of $NA_{exc}$ and $NA_{coll}$, and QY considered[23]. Taking $\langle ENH_{sig} \rangle = 0.75$ as a representative value for the planar reference, based on the results in Figure 7(b), the proposed nanowire with a taper length of $T_L = 8000$ nm achieves signal enhancements that are 2400×, 1600×, and 900× larger than the planar counterpart for $QY_0 \rightarrow 0$, 0.33, and 1, respectively. These results demonstrate that the tapered nanowire can provide more than a 2000-fold signal enhancement for low-quantum-yield emitters and nearly a 1000-fold enhancement even for emitters with unity quantum yield, highlighting its strong potential for highly sensitive fluorescence-based biosensing.

From the above $(A + B \times T_L) \times (C + D \times T_L) \times E$ for $QY_0 \rightarrow 0$, we extrapolate that a tapering length of $T_L = 19000$ nm is needed to reach a signal enhancement by a factor of 10000 compared to the corresponding planar biosensor. There, the excitation enhancement contributes 107, the collection enhancement 28, and the QY enhancement 2.5. This behavior highlights the potential of the nanowire-based platform for simple low-NA optical readout systems, while also showing that tapering offers an effective route to further boost signal enhancement.

The case of high NA ($NA_{exc} = NA_{coll} = 1$) is presented in Figure 7(c). A comparison between Figures 7(a) and 7(c) reveals that the signal enhancement at $NA_{exc} = NA_{coll} = 0.1$ is more than eight times higher than that at $NA_{exc} = NA_{coll} = 1$ for all QY cases.

From Figures 4(c) and 5(a), we observed that at higher NA, the excitation and collection enhancements vary minimally with changes in tapering length. Consequently, the signal enhancement as a function of tapering length is then almost constant at high NA, as illustrated in Figure 7(d). Thus, when comparing Figure 7(b) with Figure 7(d), it is clear that the tapering is beneficial mostly when in low-NA configuration. When we study the mixed-NA configuration with one NA low and the other NA high ($NA_{exc} = 0.1$ and $NA_{coll} = 1$; and $NA_{exc} = 1$ and $NA_{coll} = 0.1$), the signal enhancement exhibits a linear increase with increasing tapering length (Figure S11 in the Supporting Information). Such linear dependence is expected since at low NA, one of excitation or collection enhancement shows a linear increase with tapering length. Signal enhancement for the bottom-tapered configuration for these mixed NA-configuration for $NA_{exc}$ and $NA_{coll}$ is shown in Figure S17 in the Supporting Information, exhibiting similar behavior as for the top-tapered configuration above.

As expected, the proposed tapering approach is also effective for enhancing the signal in GaP nanowires. To illustrate this, we have examined top-tapered GaP nanowires, and the corresponding signal enhancement factors are presented in Section 2 of the Supporting Information.

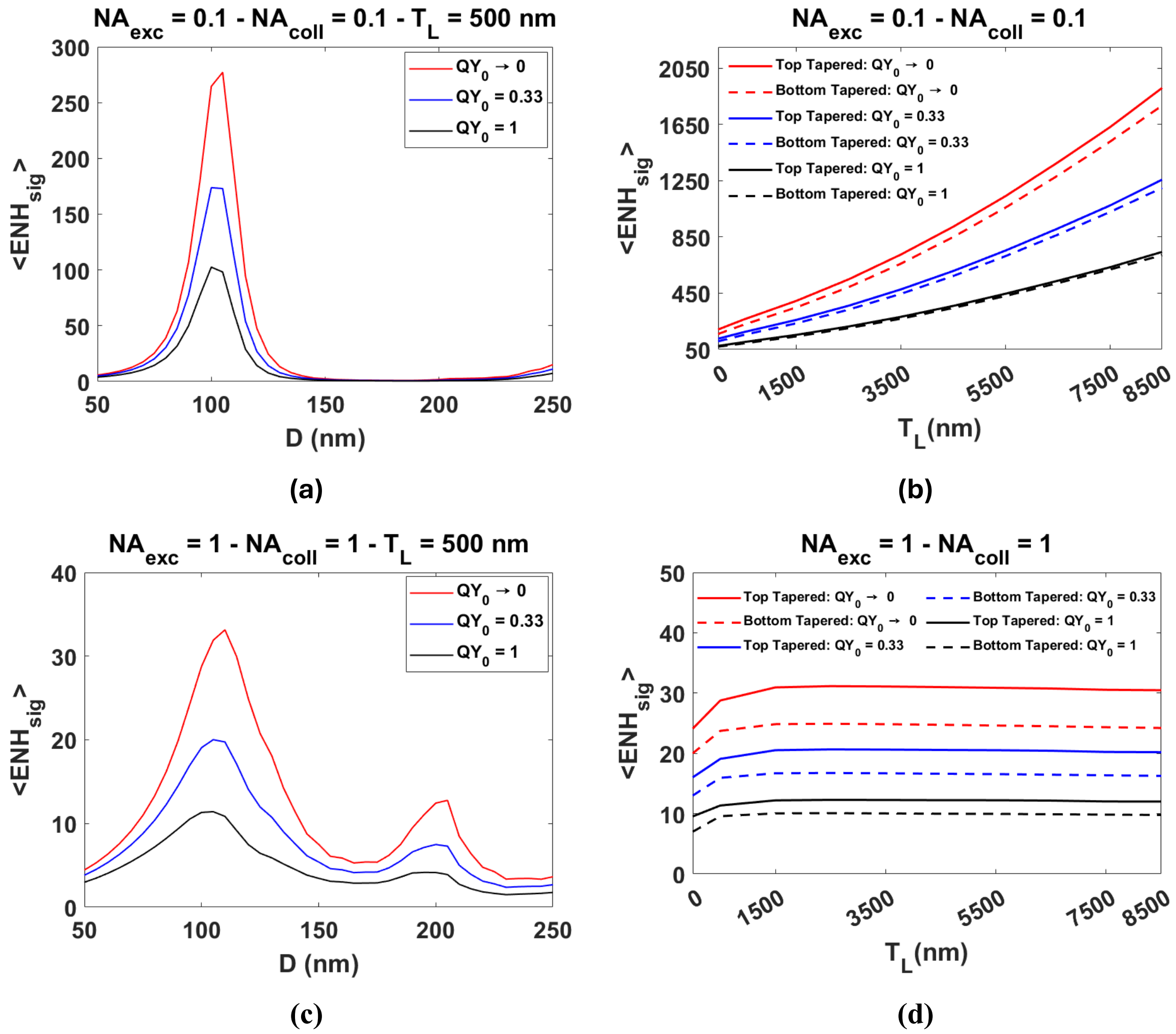

Figure 7: (a) ⟨ENHsig⟩ as a function of nanowire diameter for $QY_0$ of 0, 0.33, and 1 at $NA_{exc} = NA_{coll} = 0.1$ for top-tapered nanowire. (b) ⟨ENHsig⟩ for $QY_0 \rightarrow 0$, 0.33, and 1 for different tapering length at $NA_{exc} = NA_{coll} = 0.1$ for both top- and bottom-tapered configuration. (c) Same as (a) but for $NA_{exc} = NA_{coll} = 1$. (d) Same as (b) but for $NA_{exc} = NA_{coll} = 1$.

### 3.1 (e): Array Effects

Using an array of nanowires instead of a single nanowire could increase sensitivity and produce stronger, more measurable signals because the larger combined surface area allows a larger probability for a biomolecule to bind[30]. Furthermore, an ordered array of nanowires, with known array pattern, allows further image analysis to enhance bio-detection by combining brightfield and fluorescence microscopy in image analysis[31]. The array effect of a nanowire biosensor is important to study since multiple nanowires interact differently than a single nanowire. Elements such as the spacing between nanowires, their arrangement, and interactions like excitation shadowing and scattering of fluorescence from neighboring nanowires can significantly influence biosensor performance. Although arrays can boost sensitivity and signal strength, these interactions can also lead to signal saturation, crosstalk, or more noise[32]. Assessing these collective behaviors is crucial for ensuring consistent, reliable, and high-performing biosensors, which is essential for designing efficient and practical biosensors.

Considering these requirements, we evaluated how the array period affects the excitation enhancement (which is, as discussed in Section 3.1(d), together with the collection enhancement, the main factor that is increased by tapering). First, we chose the configuration $D$ = 100 nm, $C_L$ = 1500 nm, $T_L$ = 1000 nm, and $\lambda_{exc}$ = 640 nm for normally incident light (corresponding to $NA_{exc} \rightarrow 0$) and a hexagonal array with period $P$ (Figure 8 (a)). Compared to the single, isolated nanowire, the enhancement drops at a period below 1000 nm, while at larger period, a rather constant enhancement similar to that of the single nanowire is obtained.

Importantly, the array period that is needed for array effects to not limit the enhancement increases with the tapering length, as shown in Figure 8(b). This trend indicates that longer nanowires require a larger spacing to avoid the detrimental effect from neighboring nanowires. For instance, the minimum period needed is approximately 600 nm for the shortest tapering length of 1000 nm, while it increases to around 1700 nm for a tapering length of 8000 nm. The small markers shown at the right edge of Figure 8(b) represent the corresponding enhancement values for a single, isolated nanowire. For each tapering length, the enhancement obtained from the nanowire array approaches the value for the single nanowire as $P$ increases. At the largest simulated period of $P$ = 2500 nm, the enhancement has effectively converged to the corresponding single-nanowire value for all tapering lengths considered, confirming that array effects indeed become negligible at sufficiently large nanowire separations.

To understand why the period limits the excitation in this way, we have considered an energy balance limit on the incoupling of incident light to the $HE_{11}$ mode (monotonously increasing red line in Figure 8 (b)). The $HE_{11}$ mode carries energy in the direction of the nanowire axis.[33] We assume that with long enough tapering length, all the incident light couples into the mode. We can then calculate the maximum excitation of the $HE_{11}$ mode, which occurs when all the incident intensity into the hexagonal unit cell couples into the $HE_{11}$ mode (thus scaling with $P^2$). In this way, we obtain the excitation of the $HE_{11}$ mode[33] from which we can obtain the electric field enhancement due to this mode at the circumference of the nanowire[34].

Indeed, we find good agreement between this analysis of the maximum excitation of the $HE_{11}$ mode (monotonously increasing red line in Figure 8 (b)) and the maximum enhancement obtainable for a given tapering length (the other lines in Figure 8 (b)). Thus, we conclude that a longer tapering length allows coupling of light into the $HE_{11}$ mode from a larger surrounding, as long as the period is chosen large enough to not limit the possible coupling.

Above, we stated that <$ENH_{exc}$> of 100× is expected at a tapering length of 19000 nm for the single nanowire. From Figure 8 (b), we see that a period of 2000 nm or more is needed for such

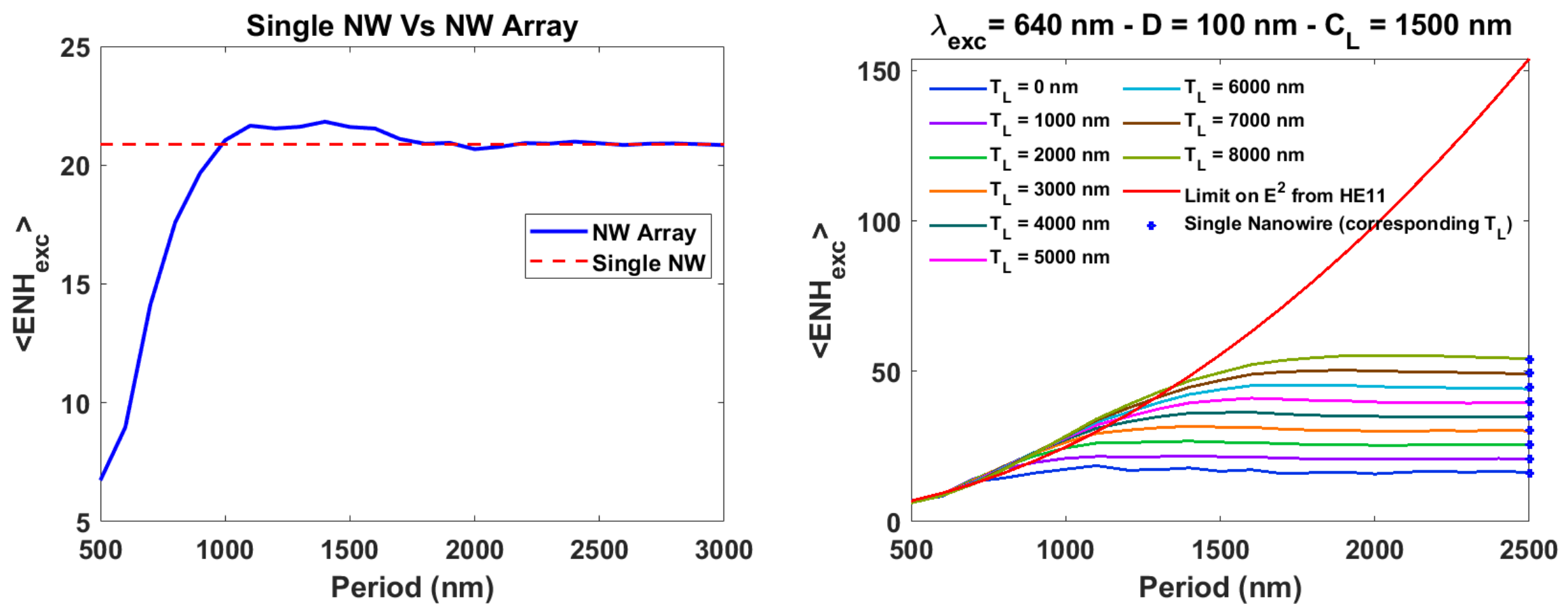

**(a)** **(b)**

Figure 8: (a) Excitation enhancement at $NA_{exc} \rightarrow 0$ of a top-tapered nanowire array with $D$ = 100 nm and $T_L$ = 1000 nm together with the enhancement for the corresponding single nanowire. (b) Same as (a) but for varying $T_L$. The monotonously increasing red line shows the limit on excitation as obtained from energy-balance analysis of the excitation of the $HE_{11}$ mode. The markers on the right edge show the corresponding value for the single nanowire case.

100× enhancement when that nanowire is placed in an array. Thus, we are looking at a rather sparse nanowire array, and one might argue that we are losing binding area for the fluorophore. However, the enhancement should not be limited by the length of the cylindrical part at which the binding occurs (since the $HE_{11}$ mode, once excited, continues propagating through the whole nanowire length). Thus, to compensate for the drop in binding area when using a large period to enable the maximum $\langle ENH_{exc} \rangle$ with a long tapering length, we can in principle increase the length of the cylindrical section proportionally to $P^2$.

### 3.2. Imaging-Mode: Signal Model and Detection Approach

We have assessed the imaging performance of the proposed tapered nanowire by simulating the imaging process (Figure 1(e)). In our design, which features a fixed cylindrical base and a tunable tapered section (Figure 2), we assess imaging from top side for both top- and bottom-tapered configurations. To quantify performance, we extract from the modelling the fraction of total intensity that falls in the image plane within a circular region of radius $r_{img}$ centered on the nanowire, denoted as $REL_{img}$[23]. This metric captures image sharpness and indicates how localized the emission is near the nanowire position in the image.

We have considered $r_{img} = 500$ nm, $\lambda_{exc} = 640$ nm, $\lambda_{em} = 670$ nm, and optimized $D = 105$ nm when using $NA_{coll} = 1$ for the imaging. For top-side imaging, $\langle REL_{img} \rangle$ decreases with increasing tapering length when we considered top-tapered geometry. This behaviour is expected as an increasing $T_L$ promotes more directional emission (see Figure S12 in the Supporting Information), corresponding effectively to a lower NA in imaging and hence less sharp image. Specifically, $\langle REL_{img} \rangle$ drops from 0.74 for $T_L = 500$ nm taper to 0.50 for $T_L = 8000$ nm (Figure 9), while the optimum focal plane shifts from approximately $z_{focal} = 1300$ nm to $z_{focal} = 1800$ nm (Figure 10).

In contrast, for the bottom-tapered nanowire with the same geometrical parameters as the above top-tapered nanowire, top-side imaging yields a virtually constant $\langle REL_{img} \rangle$ of approximately 0.83 for the full range $500 \leq T_L \leq 8000$ nm studied, while $z_{focal}$ remains essentially fixed at 1100 nm (Figure 9 and Figure 10). These results indicate that the cylindrical section retains the imaging performance of the cylindrical nanowire[23], while the tapered section continues to enhance the fluorescence signal, thanks to excitation enhancement (when exciting toward the tapered end, that is, from the bottom side through the substrate), with increasing tapering length.

It is worth noting that the bottom-tapered nanowire with top-side imaging exhibits an approximately 20% higher $\langle REL_{img} \rangle$ than the non-tapered nanowire on Si substrate (we assign this enhanced image sharpness to less of spurious reflection from the $n = 1.5$ substrate in the bottom-tapered configuration). This improvement is particularly noteworthy because the fully cylindrical nanowire itself already provides approximately a 25-fold enhancement in $\langle REL_{img} \rangle$ compared with the planar counterpart[23].

For bottom-side imaging, both configurations perform comparably to the top-side imaging (see Figures S19 and S20 in the Supporting Information). However, imaging through the Si substrate is impractical at these wavelengths due to its strong absorption and high refractive index. Bottom-side imaging for the bottom-tapered geometry is in principle feasible, as we there consider a lower-refractive-index non-absorbing substrate, such as a plastic substrate with $n \approx 1.5$. However, we lose image sharpness when imaging from the tapered end (Figure S19 in the Supporting Information). Thus, our recommendation is to use top-side imaging of the bottom-side tapered nanowire, with bottom side excitation to benefit from the tapering-induced additional excitation enhancement.

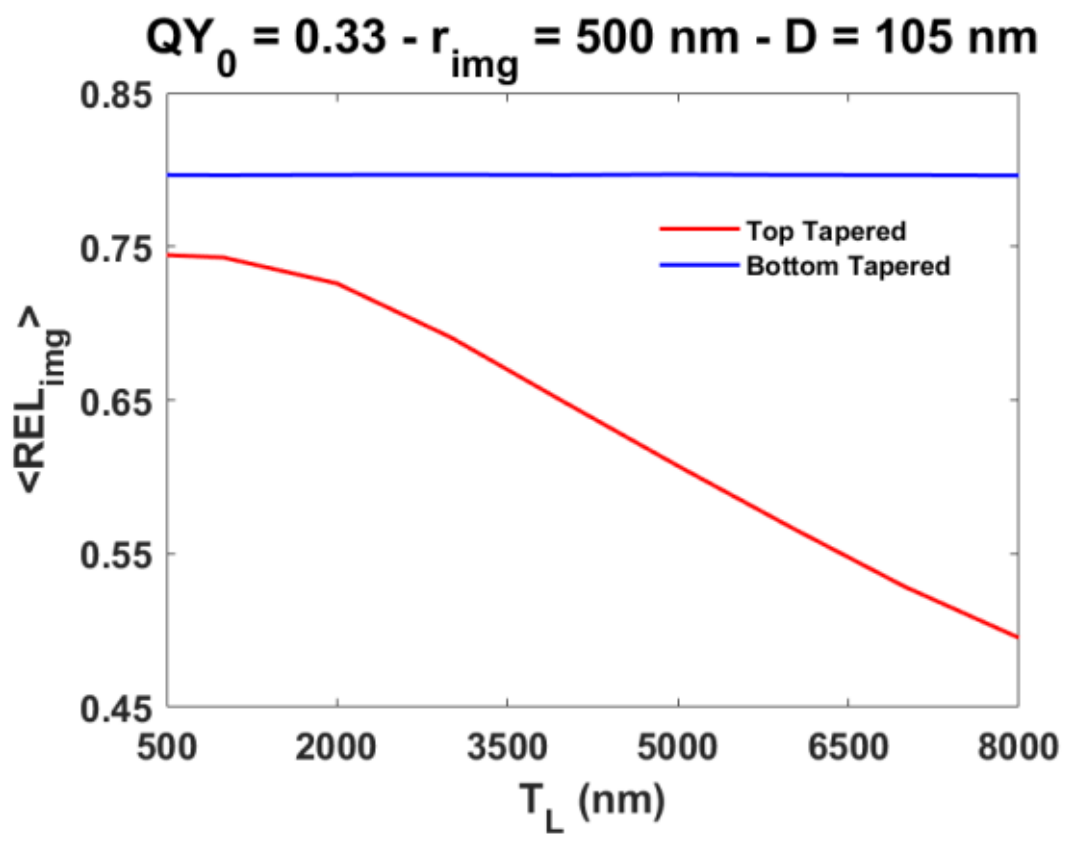

Figure 9: < $REL_{img}$ >, the integrated counts within a 500 nm radius disc in the image plane, normalized to the total counts in the image plane, for both top-tapered and bottom-tapered configurations in top-side imaging when using $NA_{coll} = 1$ for the imaging.

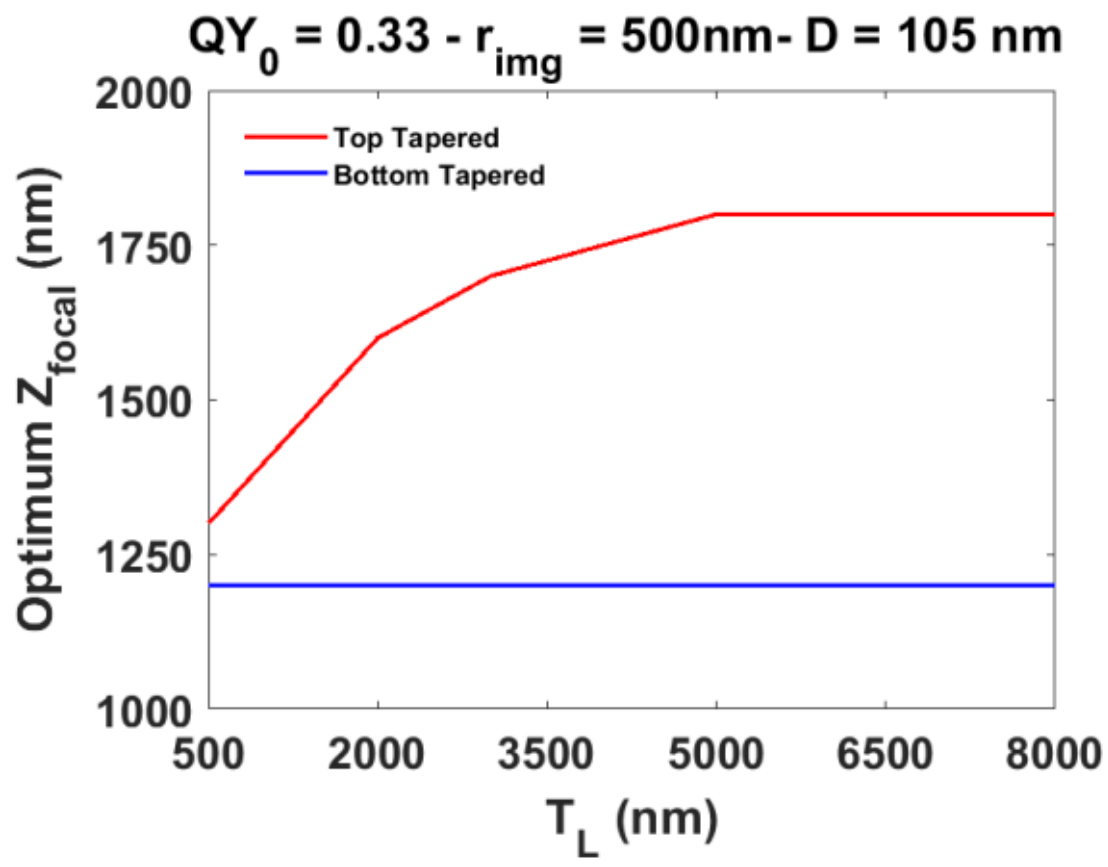

Figure 10: Optimum $z_{focal}$ as a function of tapering length for both top-tapered and bottom-tapered configurations in top-side imaging for the same system as in Figure 9.

## 4. Conclusion:

In this work, we introduced a strategically designed tapered nanowire geometry consisting of a cylindrical base and a tapered section to achieve enhanced signal detection and imaging analysis. The cylindrical section primarily supports image-sharpness in the imaging process, while the tapered section facilitates excitation enhancement by enhancing the coupling and concentration of the excitation field within the nanowire. We investigated the performance of the proposed tapered nanowire geometry under two design settings (top-tapered and bottom-tapered) using alternative approaches to address practical implementation challenges. Silicon was selected as the nanowire material due to its high refractive index, which is beneficial for enhancing the electromagnetic field and, consequently, the detected optical signal.

The nanowire diameter was kept at its optimized value, and a fixed cylindrical length was employed. Thus, only the tapering section was varied to evaluate its impact on the optical performance. All primary analysis were conducted using a single tapered nanowire to clearly understand the fundamental optical behavior. To assess practical applicability, excitation enhancement was further examined in a nanowire array configuration under optimized conditions for varying taper lengths. These results confirm that the proposed geometry remains effective in both isolated and array configurations.

We demonstrated that the taper length strongly influences fluorescence signal enhancement while having only a minimal effect on imaging performance (when performing imaging from the non-tapered end). While previous work[23] established that cylindrical nanowires significantly outperform planar fluorescence biosensors, our results show that incorporating a tapered section into the nanowire provides further gains beyond those achievable with a fully cylindrical geometry. The proposed tapered nanowire delivers more than three orders of magnitude higher fluorescence signal enhancement than the planar counterpart and consistently outperforms the non-tapered cylindrical nanowire. The fluorescence enhancement further increases with increasing taper length, and we extrapolate for a taper length of 20 μm to a $10^4$ times higher signal compared to the corresponding industry standard planar biosensor. We also demonstrate that tapered geometry can collect the same number of photons with a low numerical aperture (NA) objective as a planar biosensor operating with a high NA objective, thereby enabling simpler and more compact optical configuration for the biosensor.

In imaging mode, when imaging from the non-tapered end of the nanowire, both the top-tapered and bottom-tapered nanowire preserves the imaging performance of the cylindrical geometry, with the bottom-tapered configuration achieving even 20% higher image sharpness than a non-tapered nanowire of the same total length.

Overall, the proposed tapered nanowire enables strong signal enhancement and efficient photon collection under low NA conditions while maintaining directional emission and stable imaging performance. These characteristics make the platform well suited for cost-effective, compact, and readily integrable fluorescence biosensing applications.

**Acknowledgements**

We acknowledge the financial support of the Finnish Ministry of Education and Culture through the PREIN/I-DEEP doctoral pilot (VN/3137/2024-OKM-4) and the Research Council of Finland project 359066 (HPC-Phot). The computer resources of the Finnish IT Center for Science (CSC) and the FGCI project (Finland) are acknowledged.

# Supporting Information:

# Tapering to Enhance the Performance of Nanowire-Based Fluorescent Biosensors

Farhana Akter Mou* and Nicklas Anttu

*Physics, Faculty of Science and Engineering, Åbo Akademi University, FI-20500 Turku, Finland.*

**email: farhana.mou@abo.fi*

## Table of Contents

**Section 1: Additional technical details and electromagnetic simulation for fluorescence enhancement and imaging**.

The electromagnetic modelling framework, including the FEM formulation, dipole-emitter description, incoherent angular-spectrum treatment of wide-field excitation, Lorentz-reciprocity-based collection enhancement, Purcell factor and parasitic absorption calculations, quantum-yield modification, and fluorescence signal and imaging analysis, follows the methodology established in our previous work.[1]

In the present work, we apply this framework to a three-dimensional vertically placed tapered Silicon nanowire geometry on a substrate with an oxide coating. Both top-tapered and bottom-tapered configurations are investigated. The excitation and collection calculations are performed using full-wave 3D FEM simulations in COMSOL Multiphysics, while dipole-emission simulations are used to evaluate the Purcell factor, parasitic absorption, and far-field emission. The resulting far-field distributions are further used for fluorescence-image calculations through a near-field to far-field transformation using the RETOP package.[2]

The main modifications related to Ref. [1] concern the nanowire geometry, material composition, taper configuration, and the corresponding simulation parameters considered in this work. Unless otherwise stated, we retain the definitions, normalization procedures, dipole-orientation averaging, and modelling assumptions established in our previous work.[1] In contrast to Ref. [1] where the substrate was treated as lossless by setting the imaginary part of its refractive index to zero, the present study employs silicon as the substrate material and explicitly accounts for its complex refractive index, including the absorption-inducing imaginary component, in the electromagnetic simulations.

**Electromagnetic Simulations:**

**A. Top-tapered configuration:**

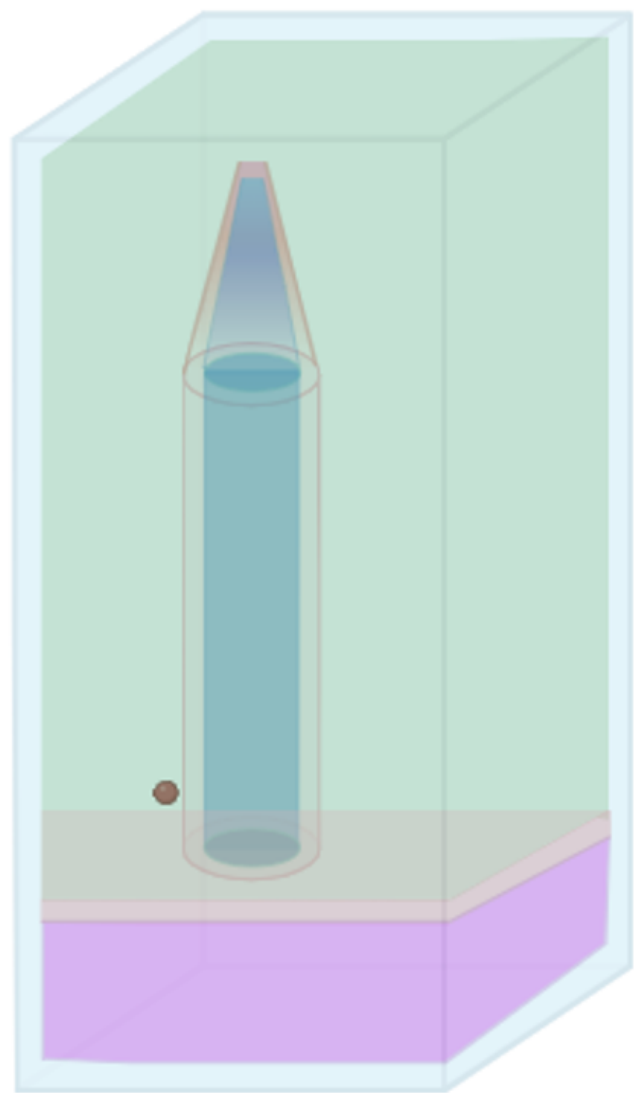

**Figure S1**: 3D view of the proposed top-tapered nanowire geometry.

## Effect of tapering on excitation enhancement at high $NA_{exc}$ = 1

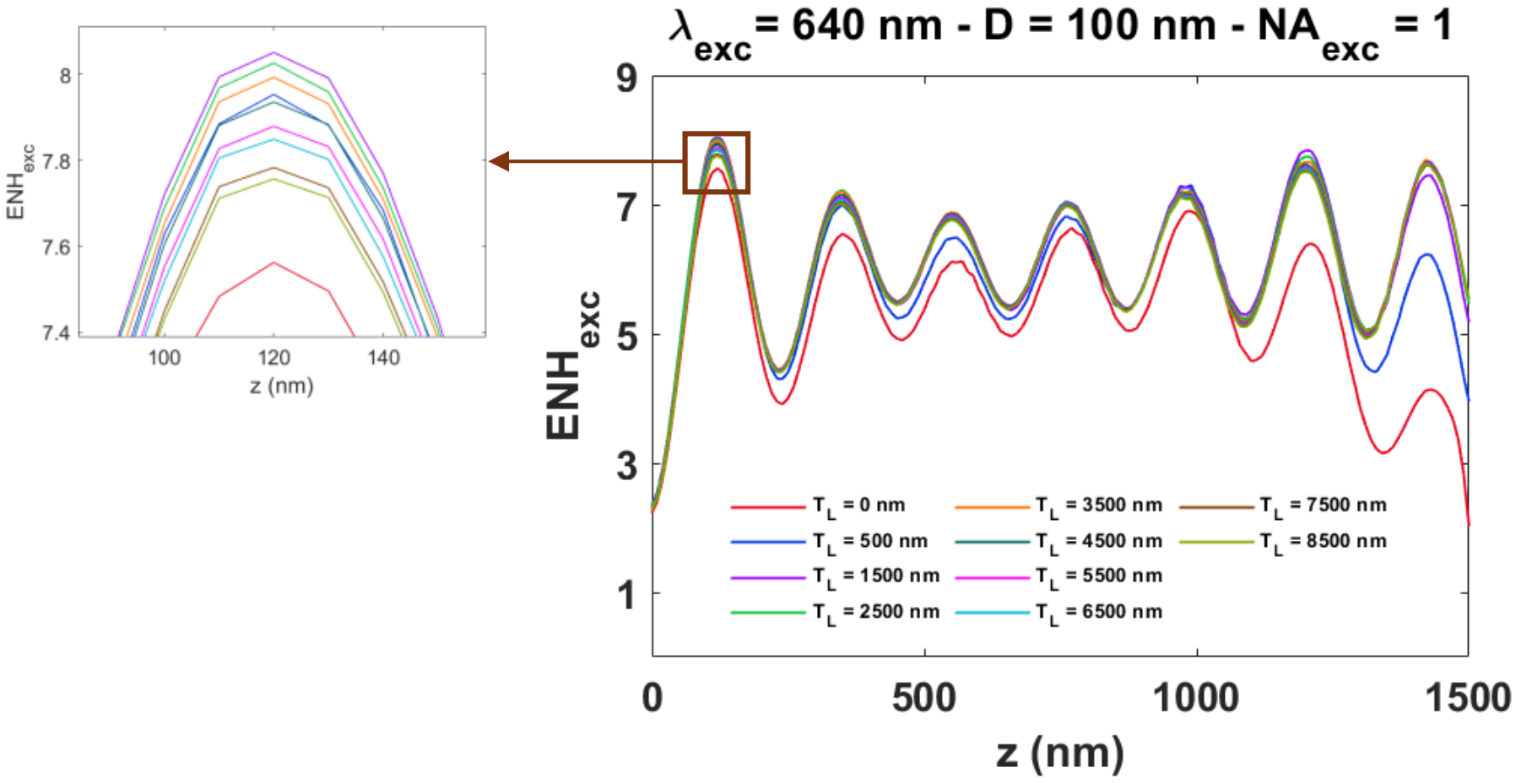

**Figure S2:** Enhancement of the incident light intensity for fluorophores located along the sidewalls of top tapered nanowire, $D$ = 100 nm, $O_t$ = 10 nm, $C_L$ = 1500 nm, and $NA_{exc}$ = 1. The variation of $EHN_{exc}$ at $\lambda_{exc}$ = 640 nm as a function of axial position where $z$ = 0 corresponds to the substrate surface and $z$ = 1500 nm denotes the top of the cylindrical binding region (i.e., the onset of the tapered section), for different taper lengths $0 \leq T_L \leq 8500$ nm. It is observed that beyond a tapering length of 2500 nm, the excitation enhancement begins to decrease. However, this change is minor, for example, decreasing only from about 8.05 at 2500 nm to 7.76 at 8500 nm.

## Effect of taper length on the electric field intensity distributions:

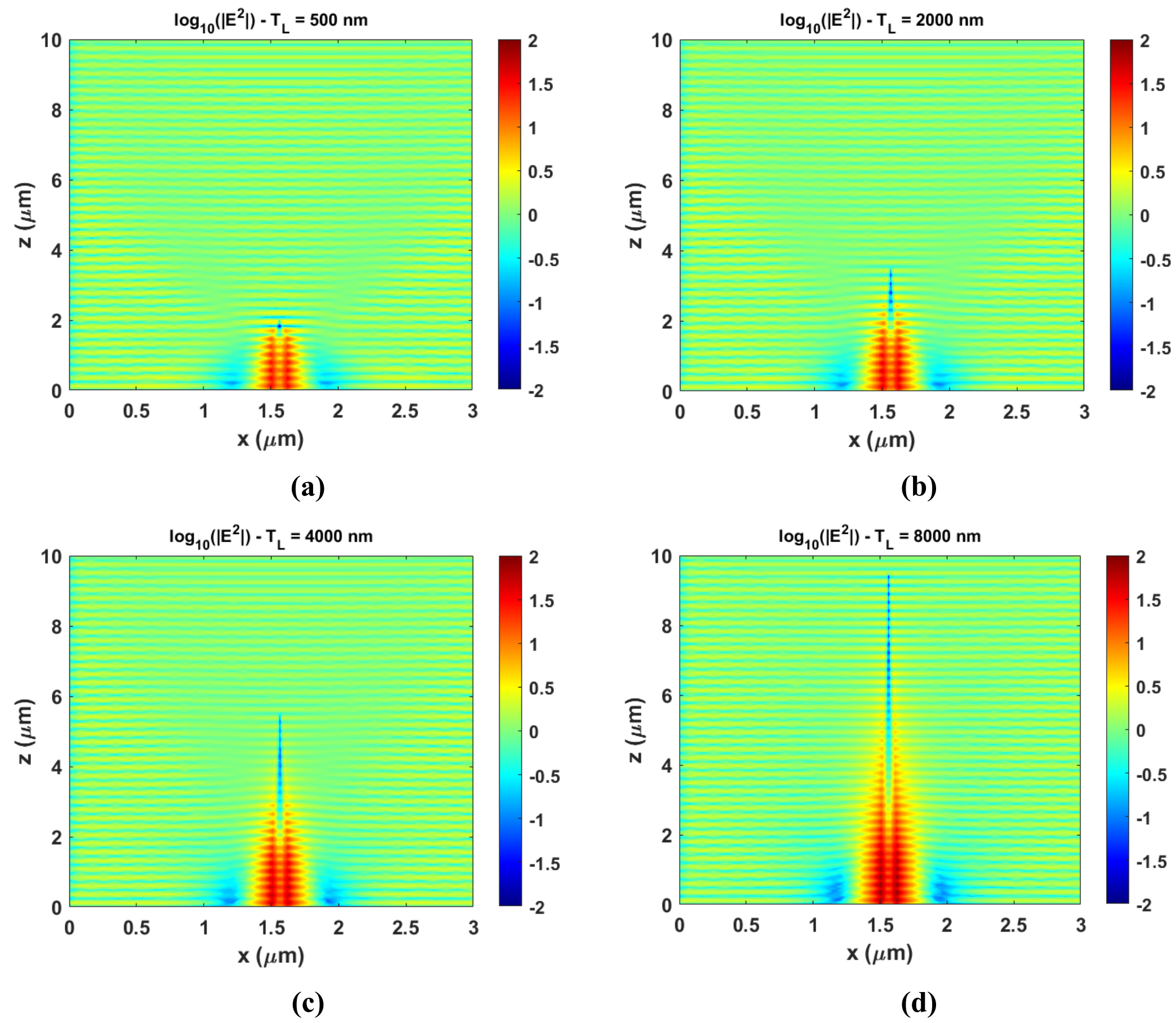

**Figure S3**: Effect of taper length on the electric field intensity distribution in the central *x*-*z* cross-section of the nanowire for polarization averaged incident light with incidence parallel to the nanowire axis. Here, $D$ = 100 nm and $\lambda_{exc}$ = 640 nm. (a) $T_L$ = 500 nm, (b) $T_L$ = 2000 nm, (c) $T_L$ = 4000 nm, and (d) $T_L$ = 8000 nm.

## Conventional ray-optics funneling and electromagnetic funneling in $HE_{11}$ guided-mode propagation:

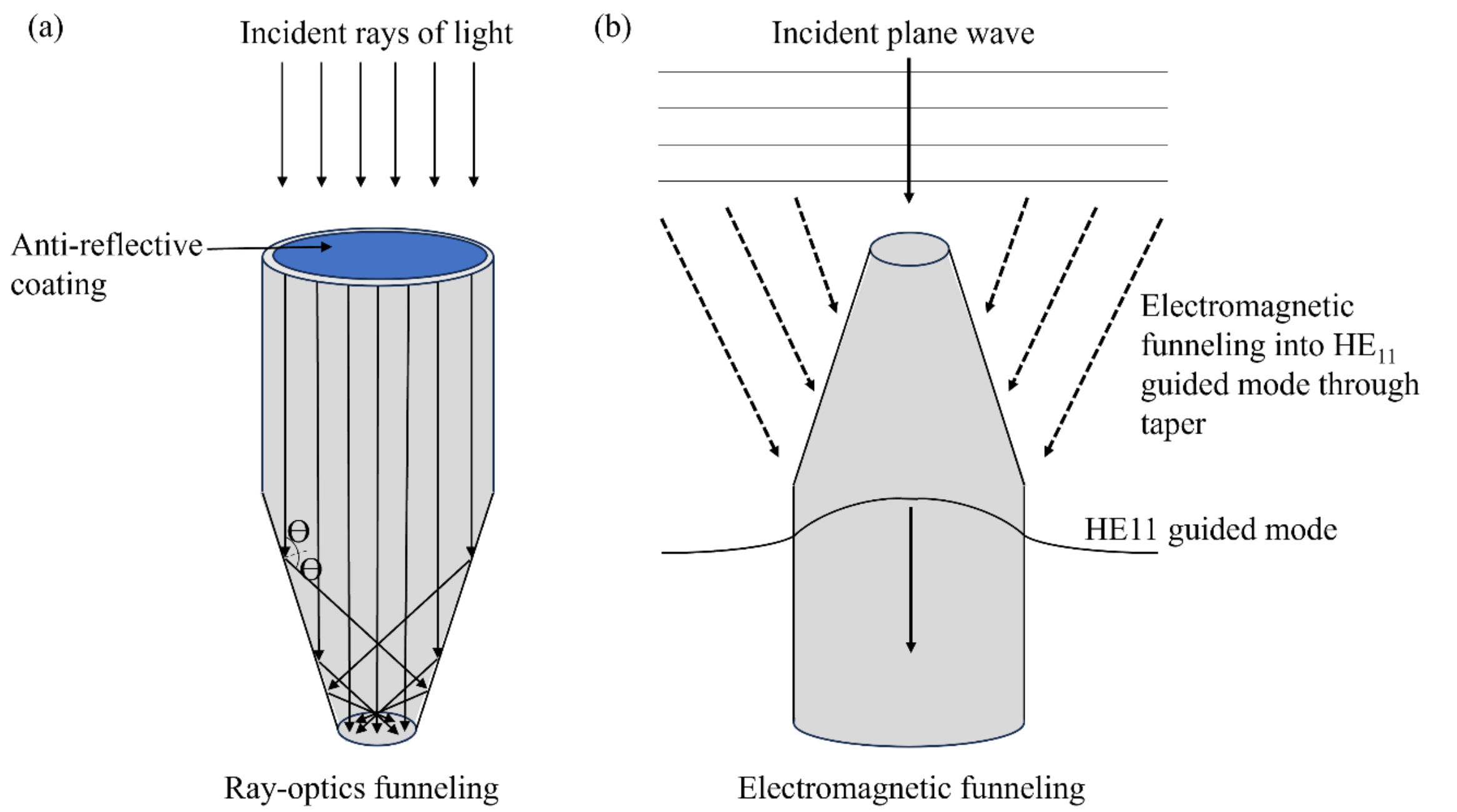

**Figure S4:** (a) Schematic of a funnel for light in the ray-optics regime. Here, rays that enter the funnel are guided through the funnel thanks to total internal reflection when the incidence angle toward the side wall of the funnel is larger than arcsin($n_2/n_1$), the angle for total internal reflection. For our case of ambient medium with $n_2$ = 1.33 and semiconductor wire with $n_1 \approx 3.5$, the angle for total internal reflection is approximately 22 degrees. The amount of light that can be funneled is limited by the geometrical top cross-section onto which the incident rays impinge. (b) Schematic of electromagnetic funneling of incident plane wave into the $HE_{11}$ guided mode of the nanowire. This type of funneling is not limited to the geometrical cross-section of the nanowire, as discussed in connection to Figure 8 in the main text. With optimized diameter, we can obtain strong field strength of the $HE_{11}$ mode at the surface of the nanowire where the fluorophores reside in our study, see Figure S3 for example for such electric field distribution.

## ⟨$ENH_{exc}$⟩: Dependence of optimum diameter on wavelength

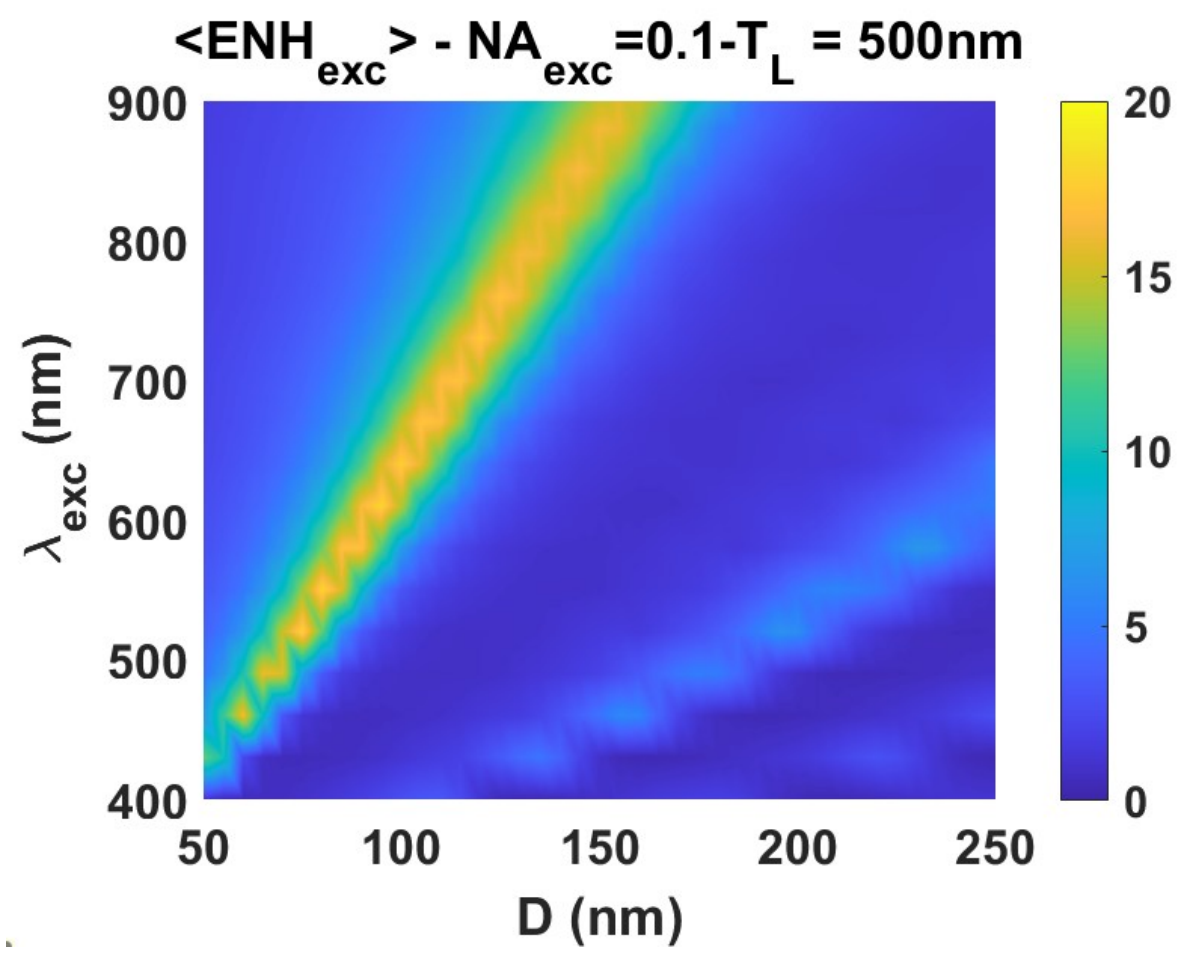

**Figure S5:** Enhancement of the incident light intensity for fluorophores located along the sidewalls of tapered Si nanowire, $C_L$ = 1500 nm, $T_L$ = 500 nm, and $O_t$ = 10 nm, as a function of $D$ for $NA_{exc}$ = 0.1 after averaging along the axial position and accounting for wavelength dependence.

### $\langle ENH_{coll} \rangle$: Dependence of collection enhancement on diameter

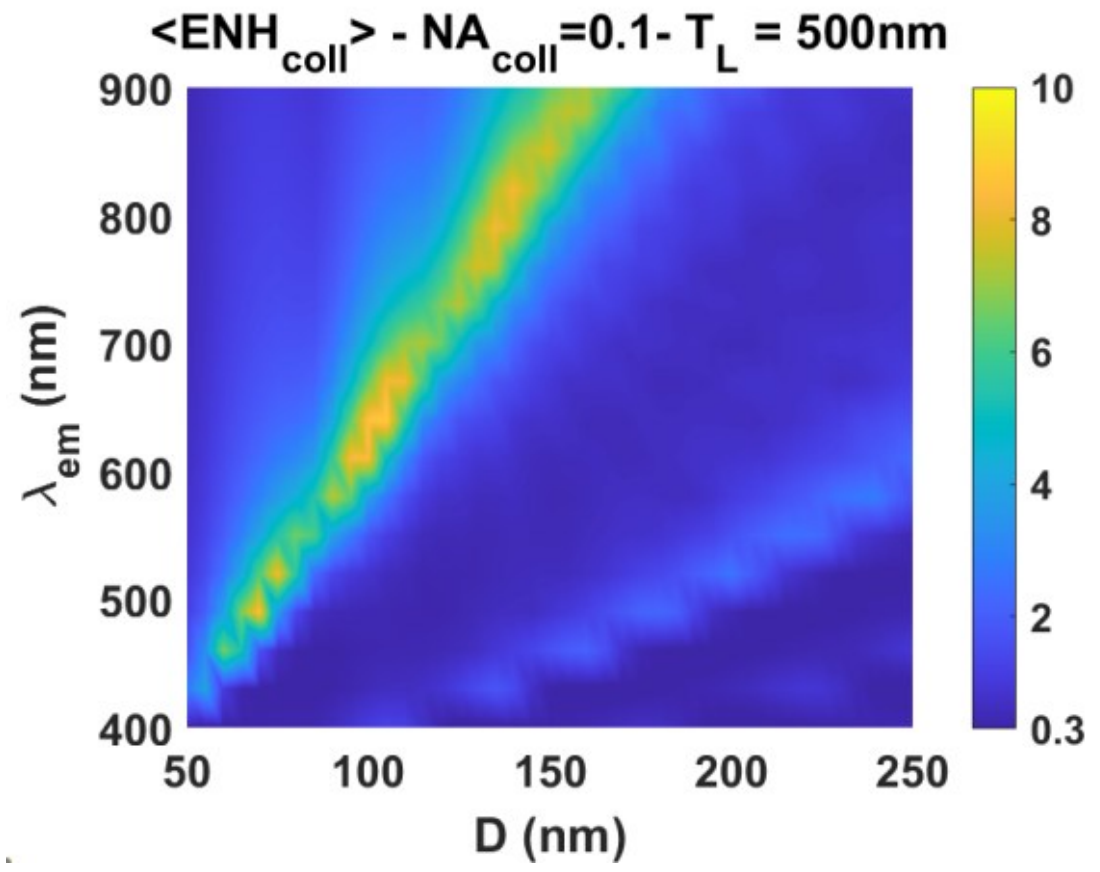

**Figure S6:** Enhancement of fluorophore emission collection at the sidewalls of top tapered nanowires for $C_L$ = 1500 nm and $O_t$ = 10 nm, as a function of emission wavelength and nanowire diameter for $NA_{coll}$ = 0.1. The result indicate that the wavelength dependence of collection enhnacement is similar to that of excitation enhancement, although the peak is slightly shifted toward shorter wavelength.

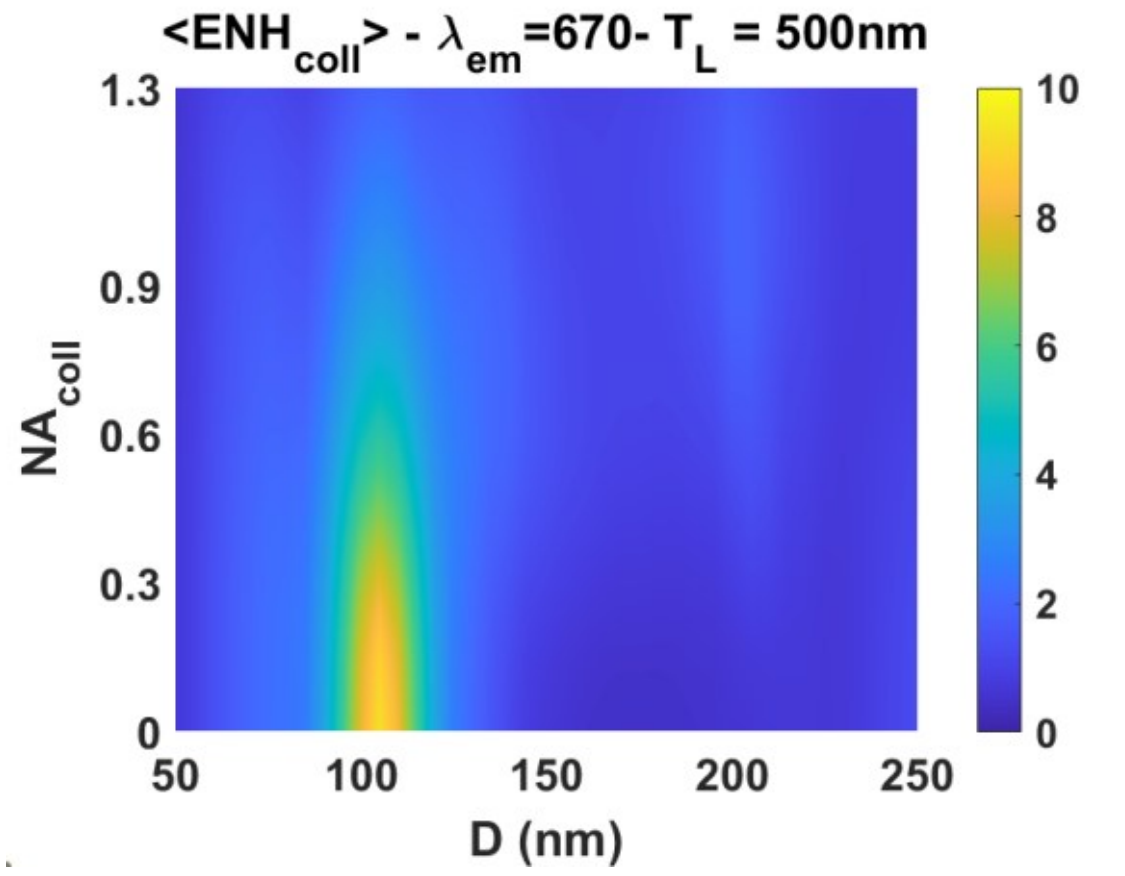

**Figure S7:** Enhancement of fluorophore emission collection at the sidewalls of top tapered nanowires for $C_L$ = 1500 nm and $O_t$ = 10 nm, as a function of $NA_{coll}$ and nanowire diameter for $\lambda_{em}$ = 670 nm.

### $<C_{Purcell}>$: Dependence of optimum diameter on wavelength

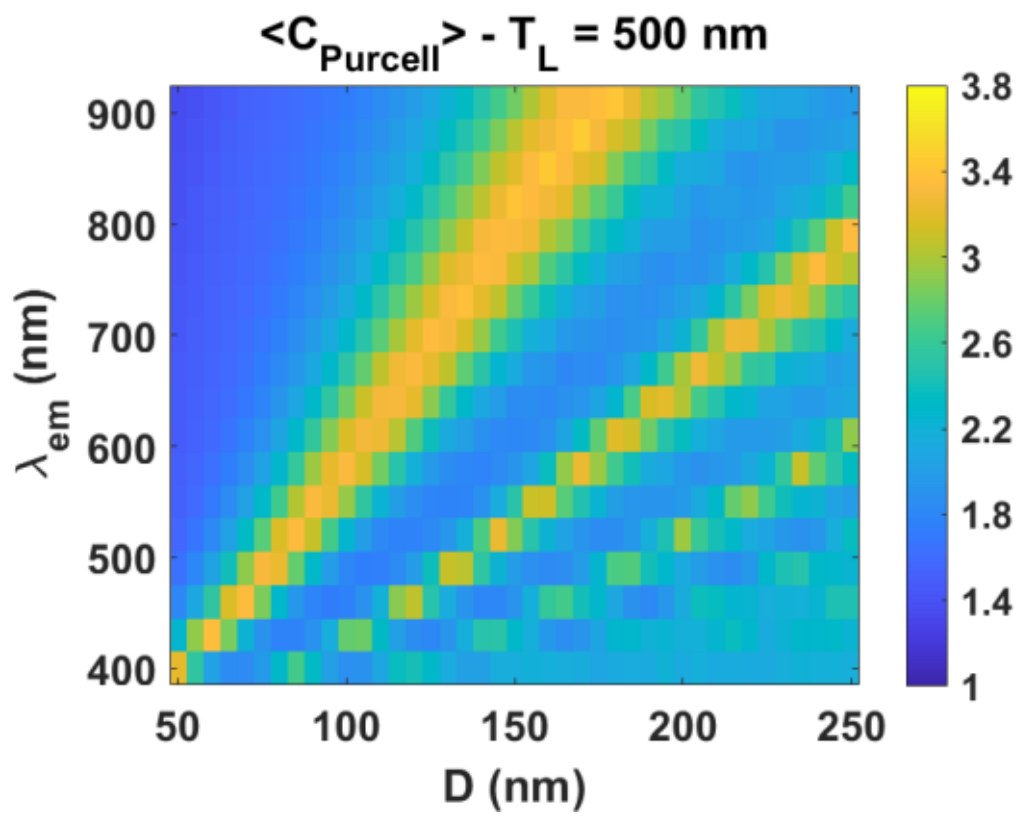

**Figure S8:** Purcell factor of top tapered nanowires for $C_L$ = 1500 nm and $T_L$ = 500 nm, as a function of emission wavelength and nanowire diameter.

### Absorption loss:

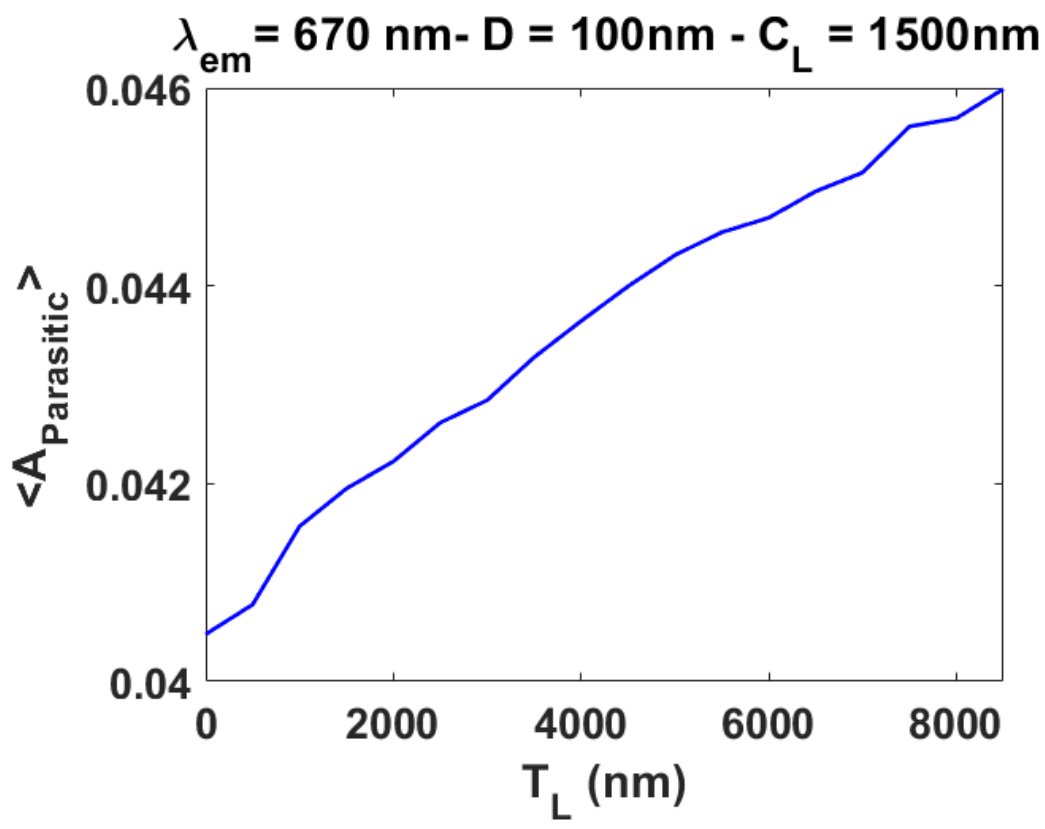

**Figure S9:** Absorption loss of emission from fluorophores bound to a Si nanowire, as a function of tapering length. The absorption loss exhibits an approximately linear dependence on taper length, but the magnitude of the change is modest, rising from 0.040 in the absence of tapering to 0.046 for an 8500 nm long taper.

## Signal Enhancement at different $NA_{exc}$ and $NA_{coll}$

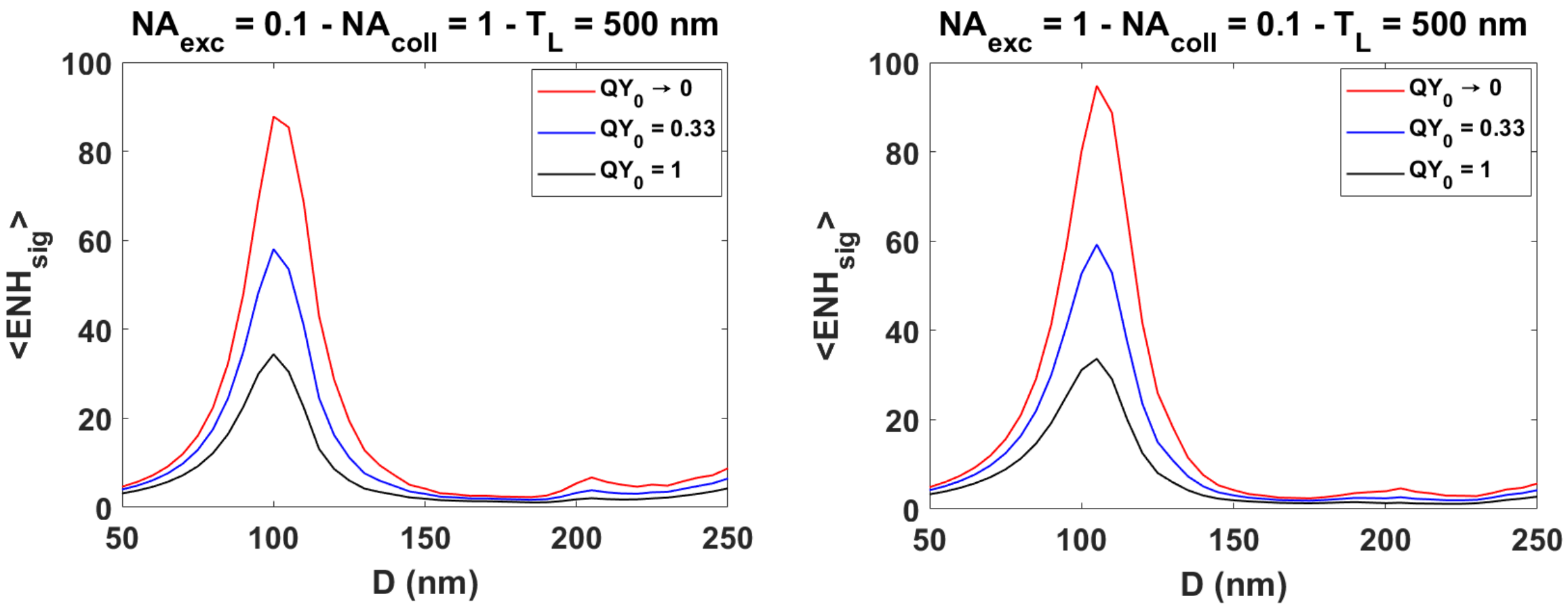

**(a)** **(b)**

**Figure S10:** The average signal enhancement ⟨ENHsig⟩ from fluorophores bound to a top-tapered Si nanowires, as a function of nanowire diameter for $QY_0 \rightarrow 0$, 0.33, and 1 at (a) $NA_{exc} = 0.1$ and $NA_{coll} = 1$; and (b) $NA_{exc} = 1$ and $NA_{coll} = 0.1$.

## ⟨ENHsig⟩: Effect of Taper Length and NA

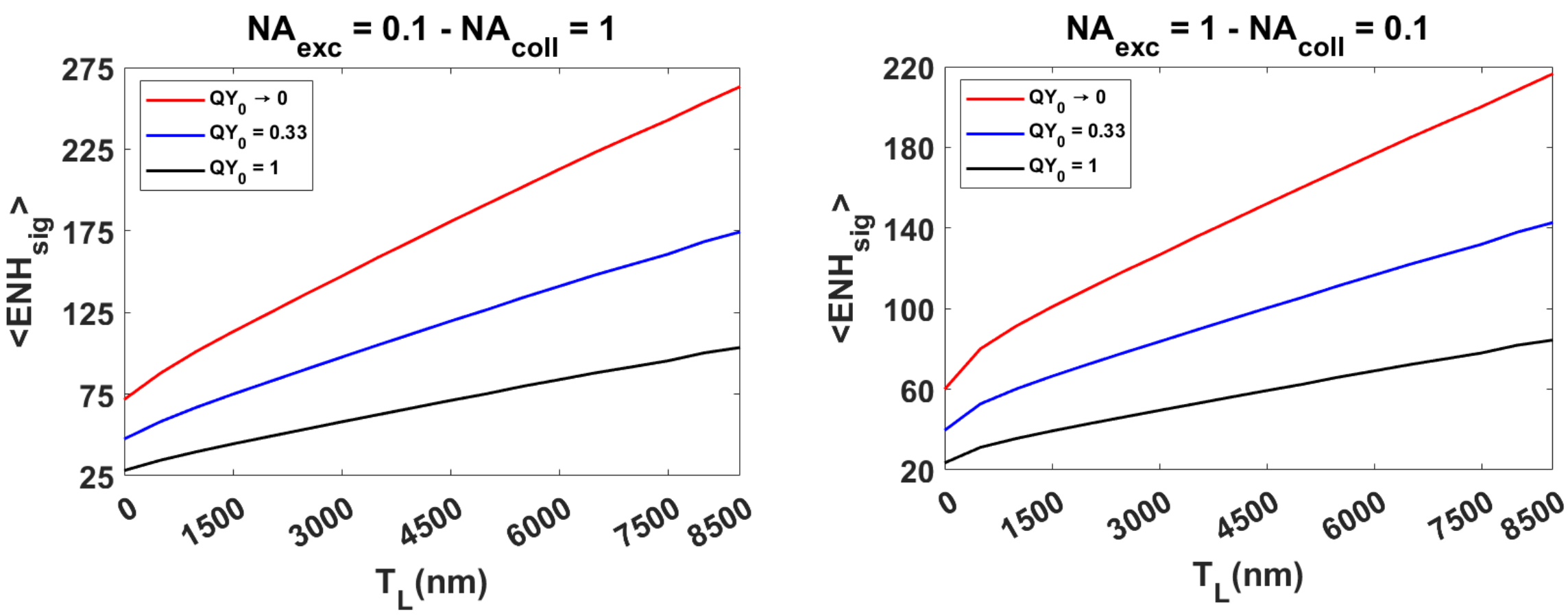

**(a)** **(b)**

**Figure S11:** The average fluorescence signal enhancement ⟨ENHsig⟩ of Si top tapered nanowires as a function of tapering length ($T_L$) for $QY_0 \rightarrow 0$, 0.33, and 1 at (a) $NA_{exc} = 0.1$ and $NA_{coll} = 1$; and (b) $NA_{exc} = 1$ and $NA_{coll} = 0.1$.

### Directional emission with taper length

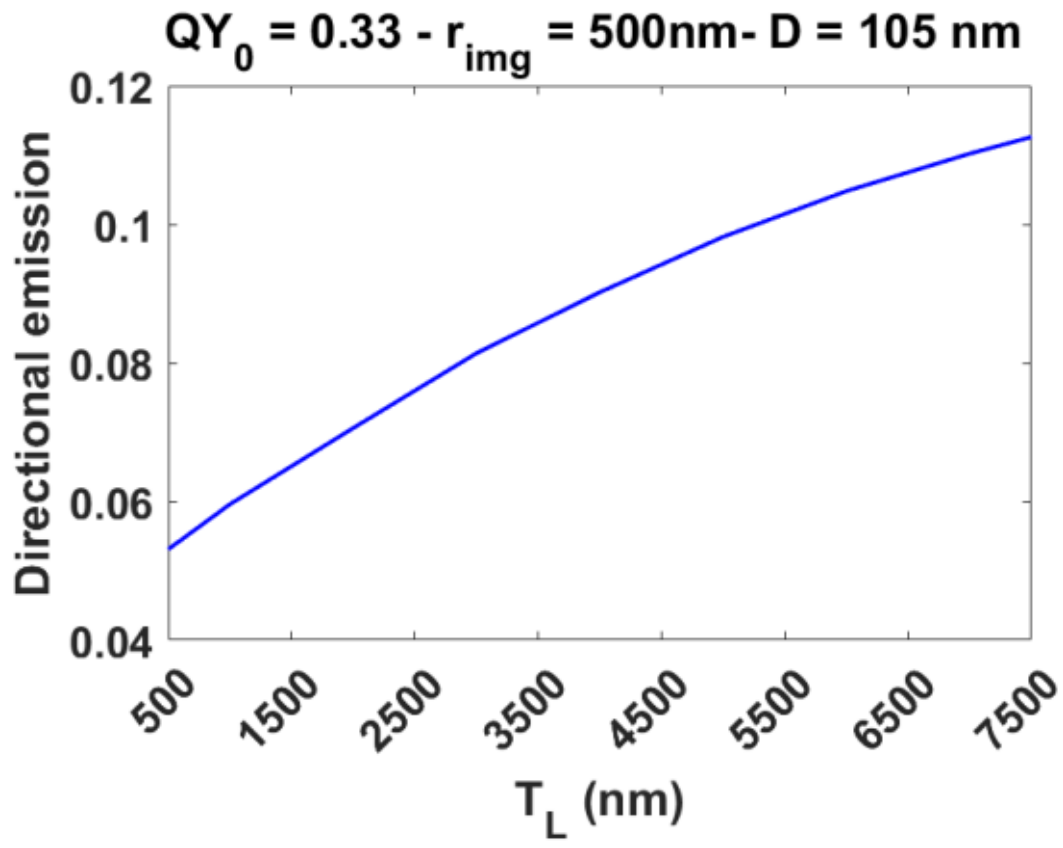

**Figure S12:** Directional emission of the top-tapered nanowire for varying tapering length ($T_L$) with $QY_0 = 0.33$, $r_{img} = 500$ nm and $D = 105$ nm. Emission directionality is defined as the fraction of total top-side radiated power emitted within a 5° half-angle around the surface normal. As the taper length increases, a larger fraction of the emitted power is concentrated near the surface normal, resulting in more than a twofold enhancement in directionality over the investigated range.

## B. Bottom Tapered Configuration:

### Geometrical setup for avoiding selective functionalization

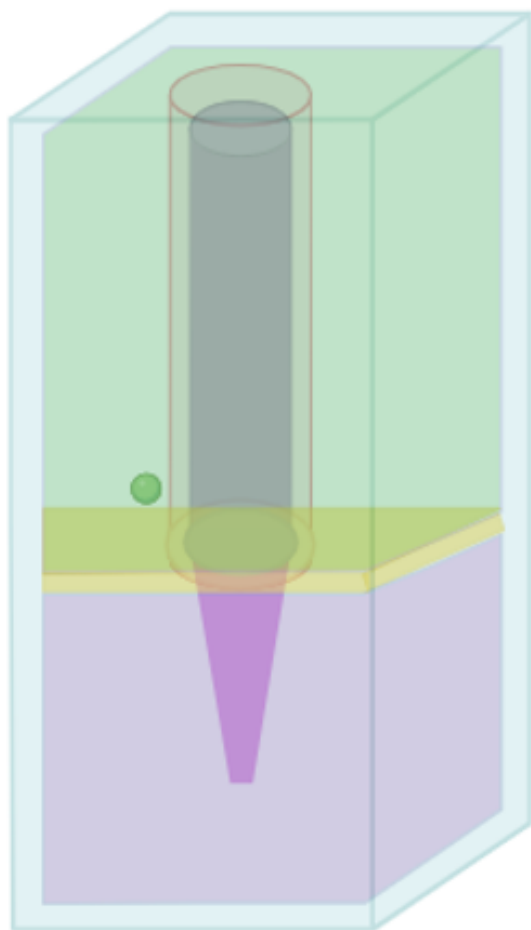

**Figure S13**: Proposed geometrical setup for avoiding selective functionalization. The tapered part is placed inside the substrate of a lower refractive index than Si, e.g. a plastic membrane with $n \approx 1.5$.

## Excitation Enhancement:

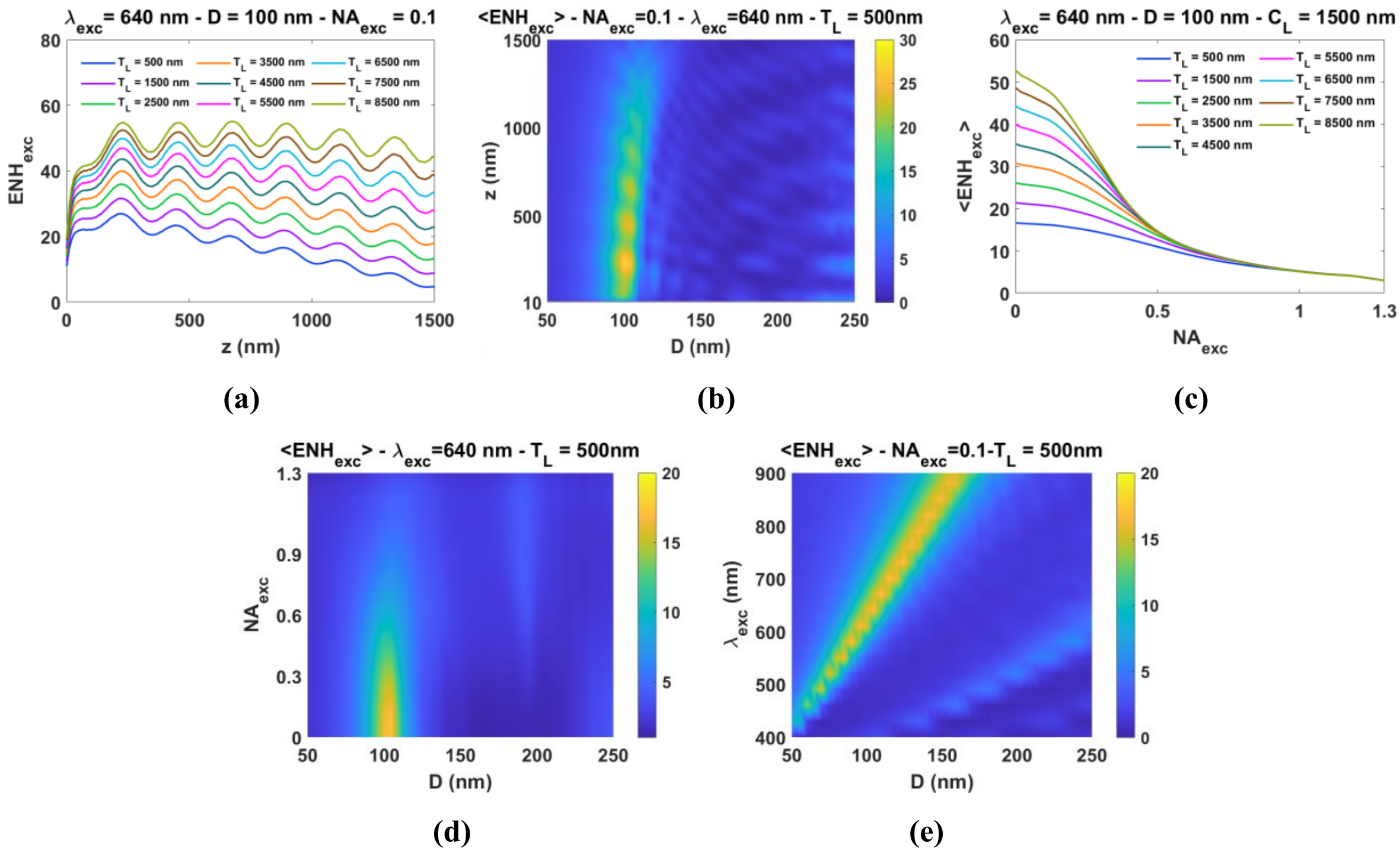

**Figure S14:** Enhancement of the incident light intensity for fluorophores located along the sidewalls of a bottom-tapered nanowire with $C_L$ = 1500 nm and $O_t$ = 10 nm. (a) $EHN_{exc}$ at $\lambda_{exc}$ = 640 nm and $D$ = 100 nm as a function of axial position with $z$ = 0 corresponding to the substrate surface and $z$ = 1500 nm denoting the top of the cylindrical binding region (i.e., the onset of the tapered section). (b) same as (a) but with diameter dependence for $T_L$ = 500 nm. (c) $\langle ENH_{exc}\rangle$ as a function of $NA_{exc}$ for $D$ = 100 nm and different tapering lengths. (d) Same as (c) but with diameter dependence for $T_L$ = 500 nm. (e) same as (d) but as a function of emission wavelength for $NA_{exc}$ = 0.1.

## Collection Enhancement:

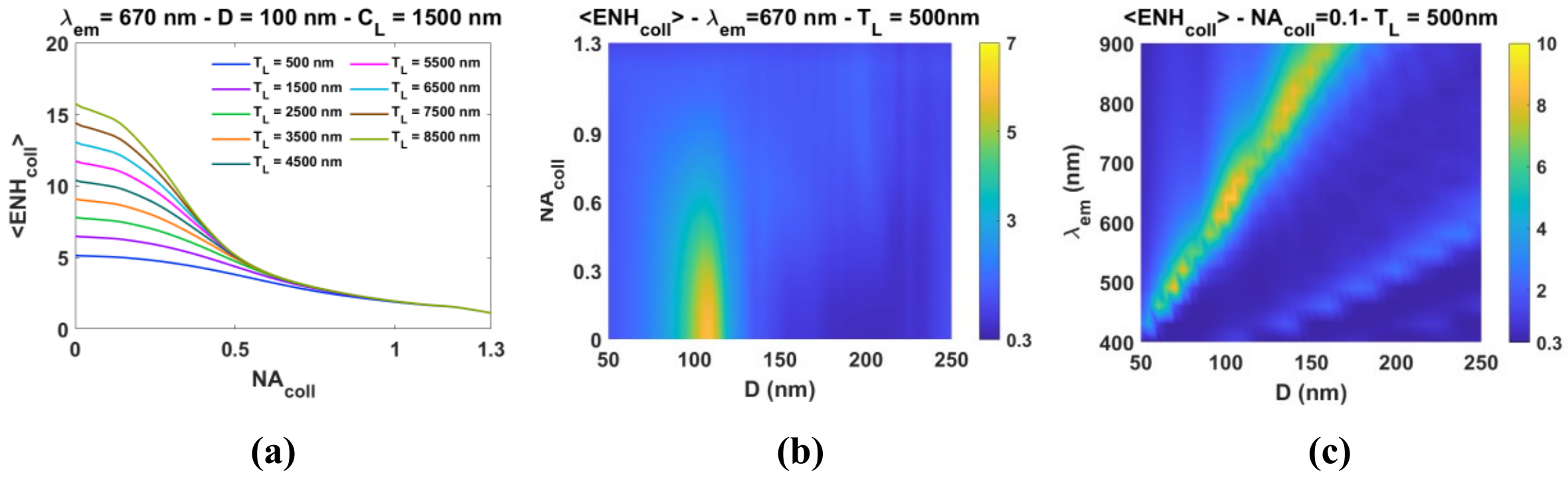

**Figure S15:** Enhancement of emission collection from fluorophores at the sidewall of a bottom-tapered nanowire for $C_L$ = 1500 nm and $O_t$ = 10 nm (a) as a function of $NA_{coll}$ for a constant emission wavelength of $\lambda_{em}$ = 670 nm for $D$ = 100 nm and different tapering length, (b) as a function of $NA_{coll}$ and nanowire diameter for $\lambda_{em}$ = 670 nm and $T_L$ = 500 nm, and (c) as a function of emission wavelength and nanowire diameter for $NA_{coll}$ = 0.1.

## Purcell factor and parasitic absorption:

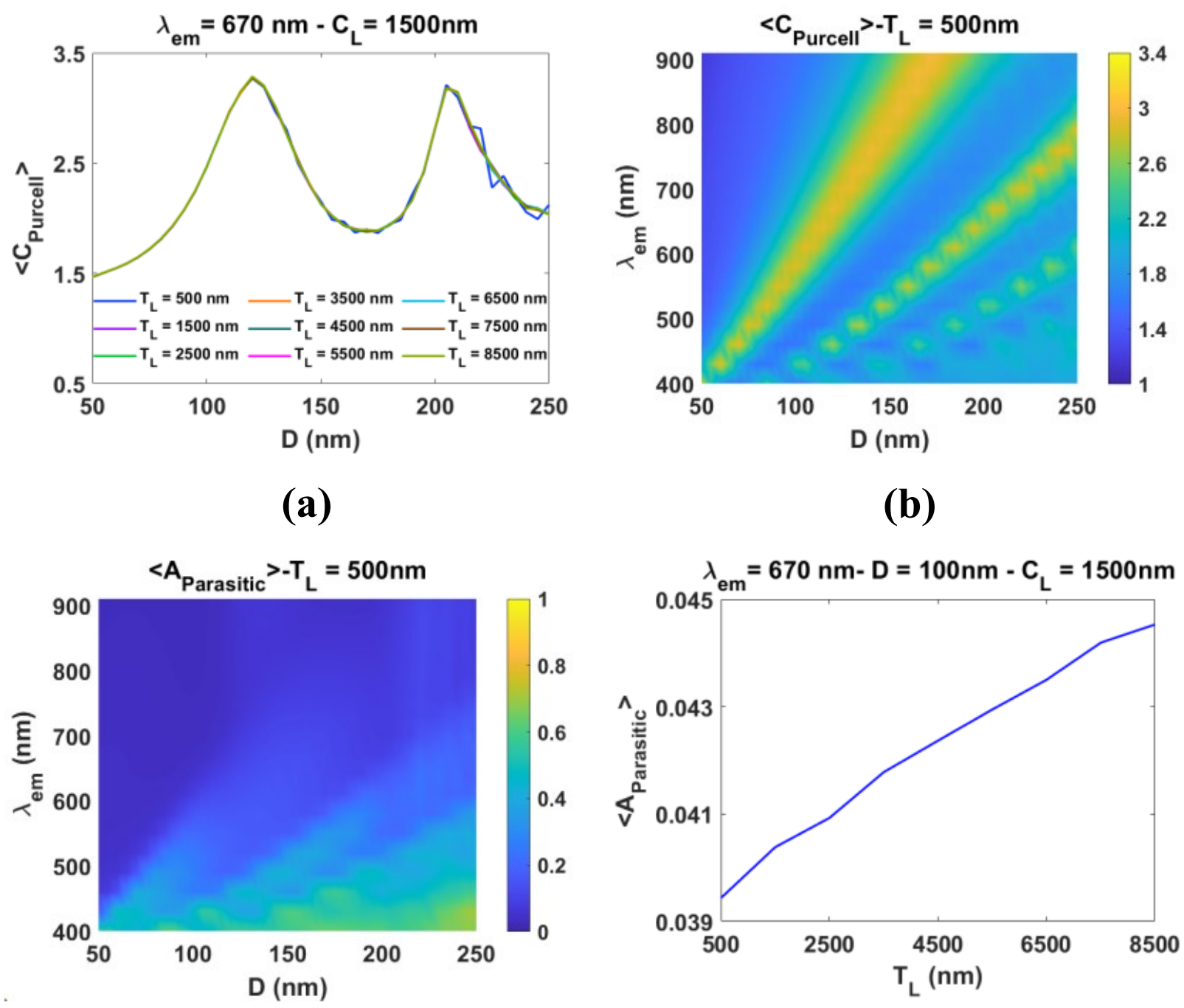

**(a)** **(b)**

**(c)** **(d)**

**Figure S16:** (a) Purcell factor for different tapering length as a function of diameter for bottom-tapered geometry. (b) Purcell factor of bottom-tapered nanowire as a function of emission wavelength and nanowire diameter. (c) Parasitic absorption of emission from fluorophores at the sidewalls of bottom-tapered nanowires for $T_L = 500$ nm, and $O_t = 10$ nm, as a function of emission wavelength and nanowire diameter. (d) Absorption loss profile with respect to tapering length at $D = 100$ nm.

## Signal Enhancement:

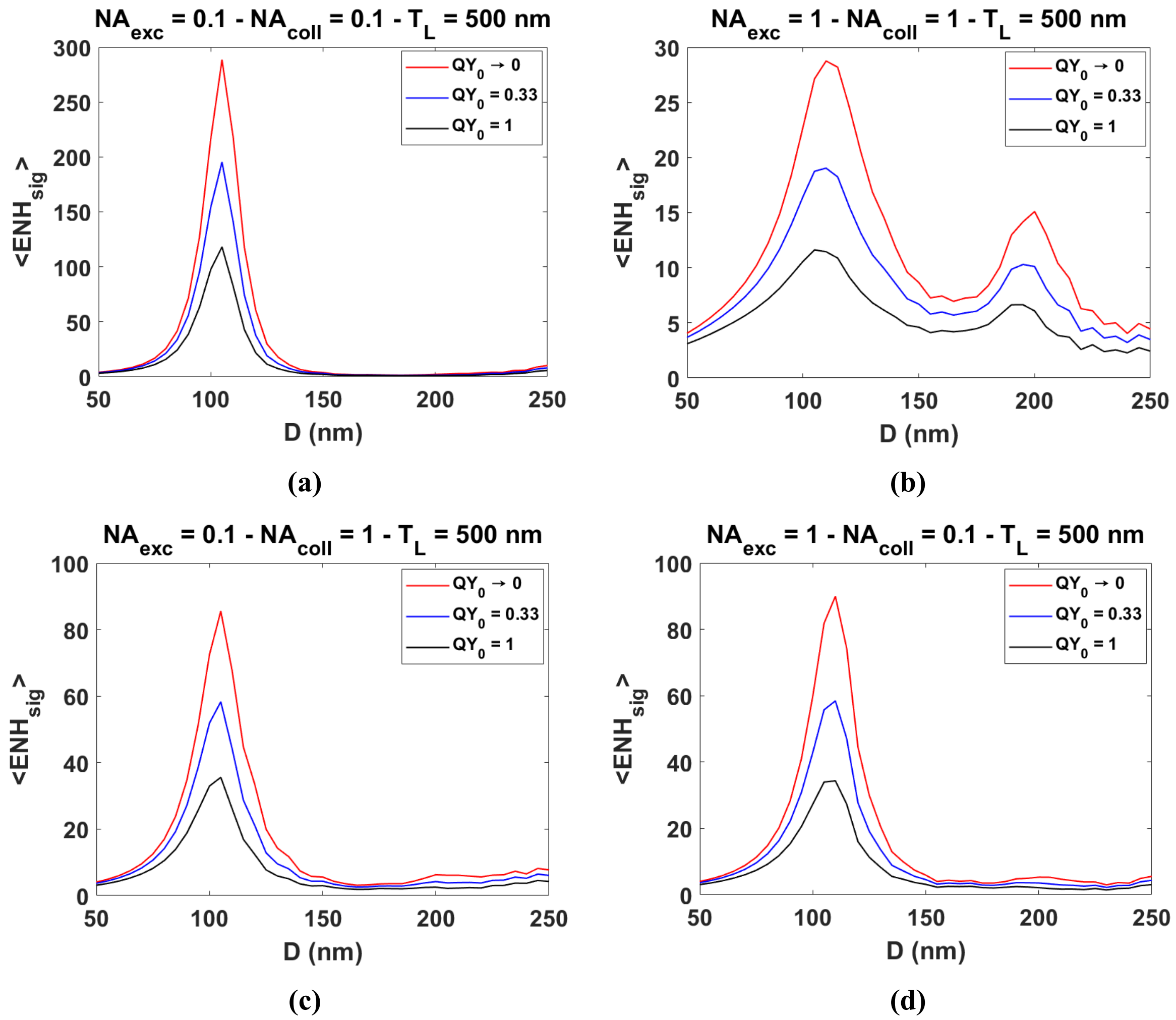

**Figure S17:** ⟨$ENH_{sig}$⟩ for bottom-tapered nanowires as a function of nanowire diameter for $QY_0 \rightarrow 0$ , 0.33, and 1 at (a) $NA_{exc} = 0.1$ and $NA_{coll} = 0.1$, (b) $NA_{exc} = 1$ and $NA_{coll} = 1$, (c) $NA_{exc} = 0.1$ and $NA_{coll} = 1$, and (d) $NA_{exc} = 1$ and $NA_{coll} = 0.1$.

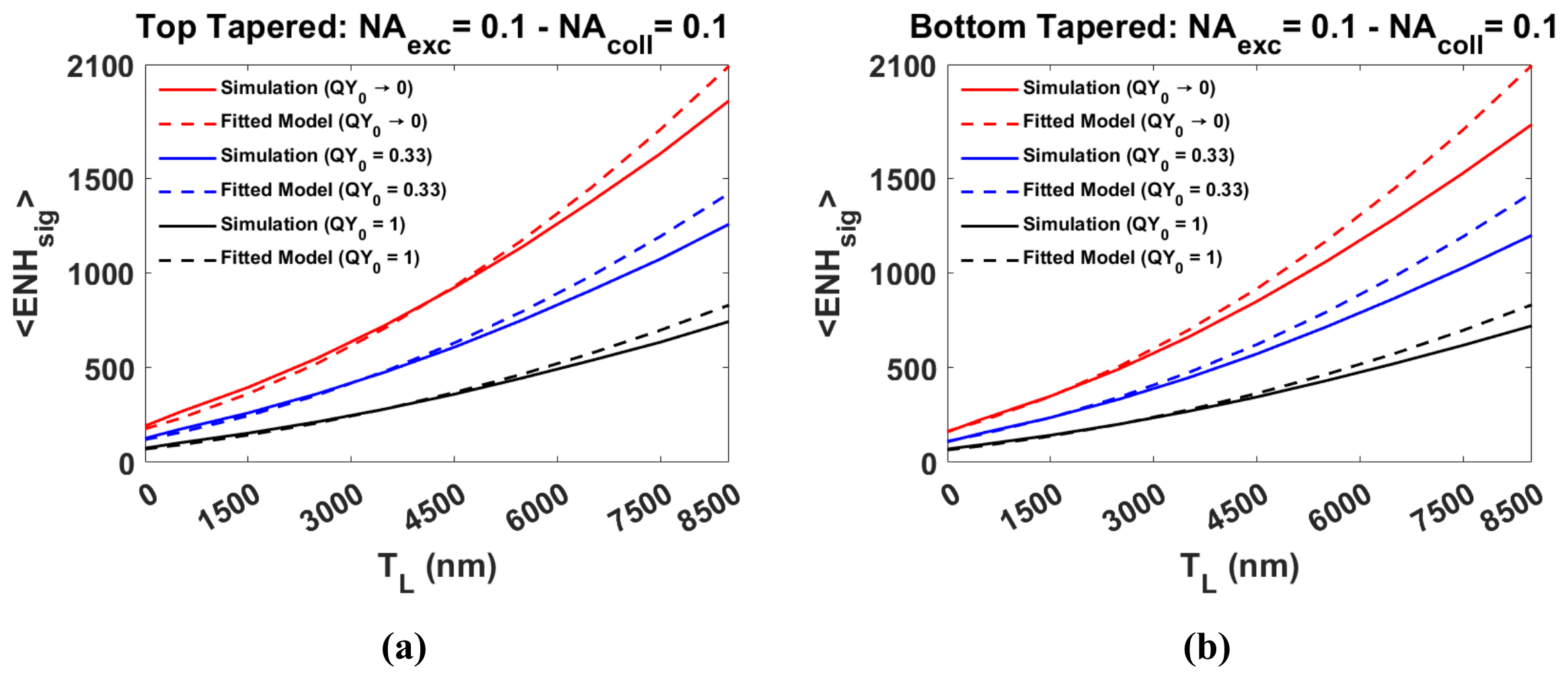

**Figure S18:** ⟨$ENH_{sig}$⟩ as a function of tapering length for $NA_{exc} = NA_{coll} = 0.1$ for both (a) top-tapered and (b) bottom-tapered nanowires, including the results from the model $(A + B \times T_L) \times (C + D \times T_L) \times E/F/G$ presented in the main text (dashed lines).

## C. Comparative Imaging Analysis

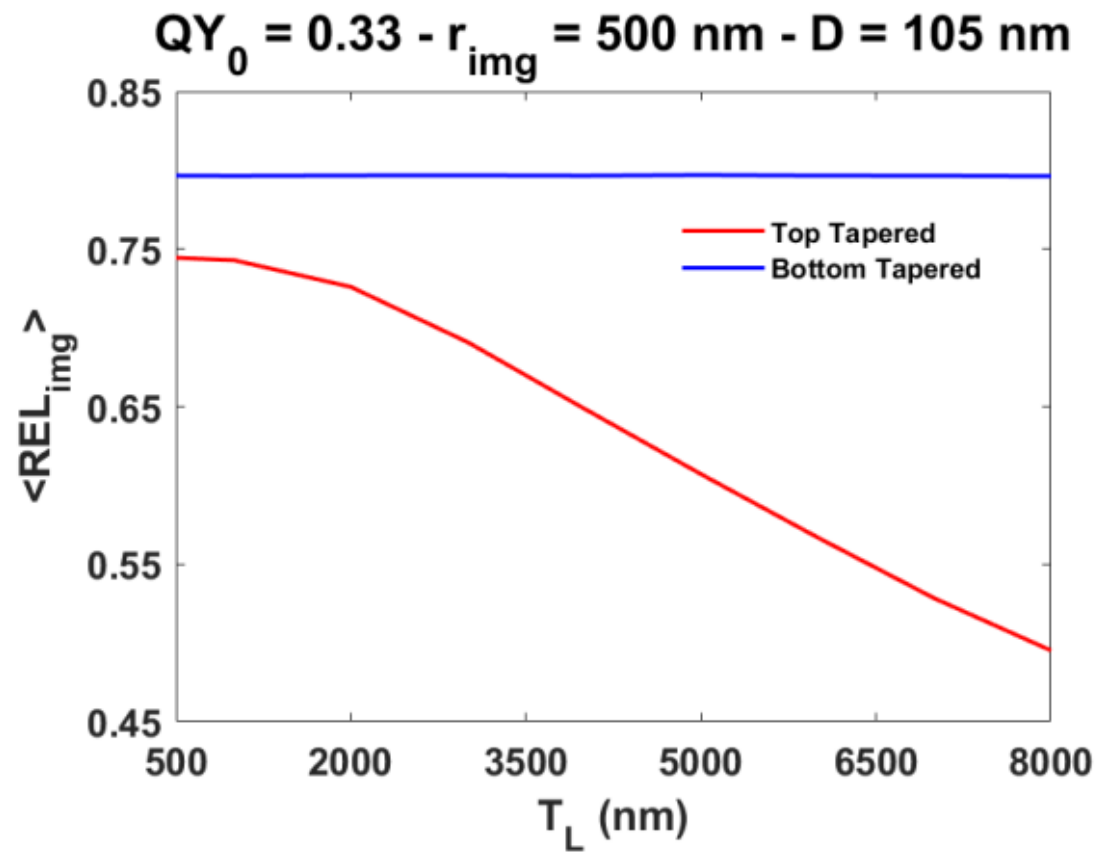

**Figure S19:** Integrated counts within a 500 nm radius disc in the image plane, normalized to the total counts in the image plane, for both the top-tapered and bottom-tapered configurations in bottom-side imaging mode when using $NA_{coll}$ = 1 for the imaging. Here, the results for the top tapered configuration do not take into account the absorption in the Si substrate.

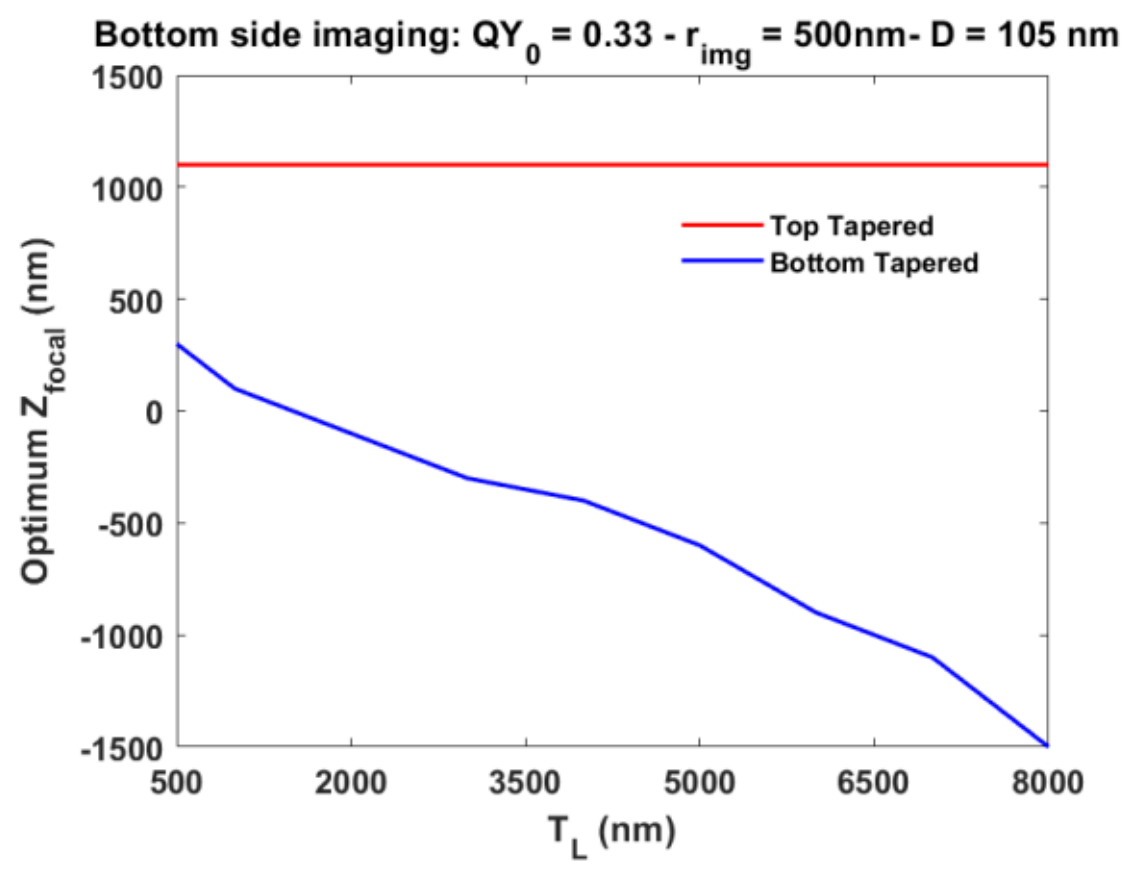

**Figure S20:** Optimum $Z_{focal}$ with the tapering length for both top and bottom tapered situations for the same case as considered in Figure S19.

## Fabrication feasibility of proposed tapered nanowire:

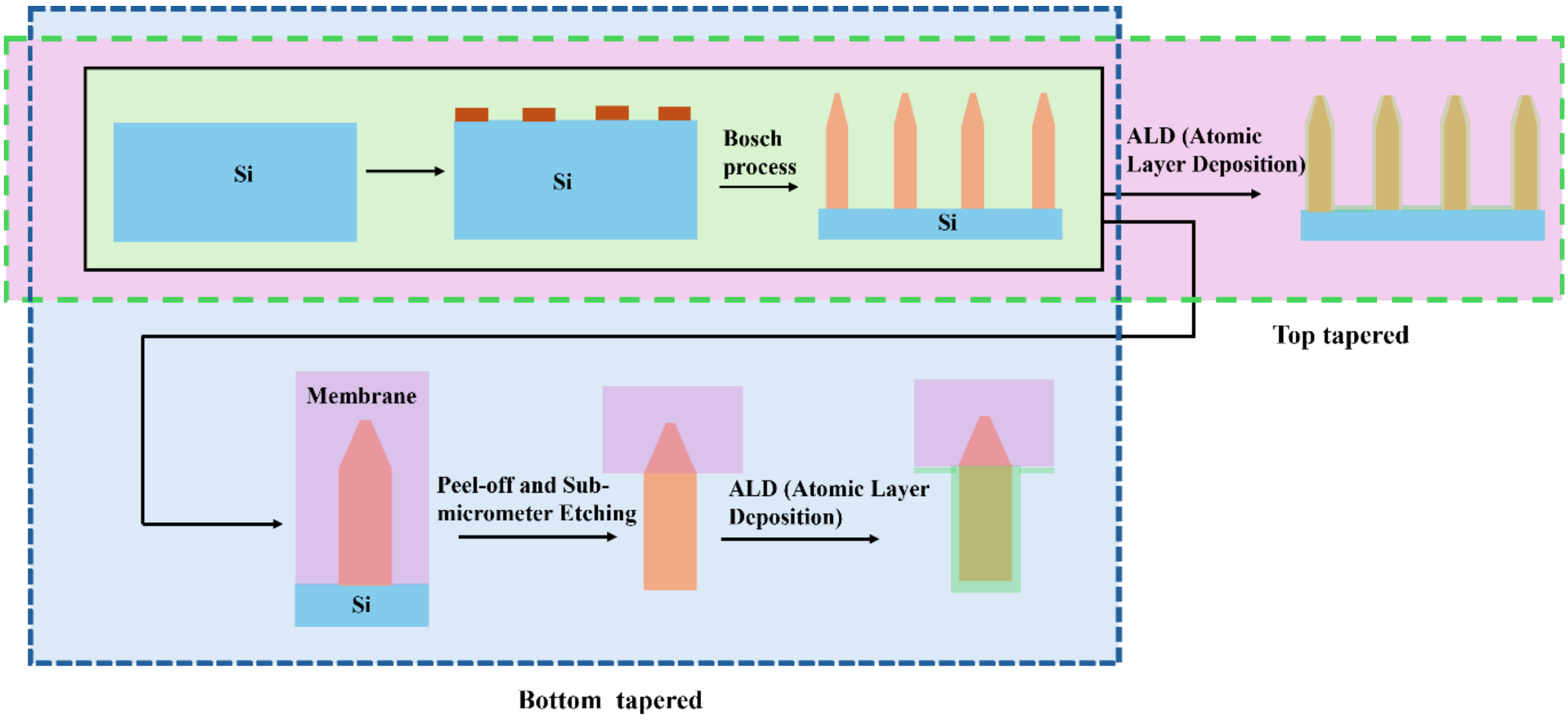

**Figure S21:** Schematic of the fabrication process of proposed tapered nanowire. A possible fabrication route for the proposed top- and bottom-tapered nanowire structures on cost-effective bulk silicon substrates could involve a top-down fabrication approach. The process could begin with standard lithographic patterning on a bulk silicon wafer, followed by deep reactive-ion etching (DRIE) using a Bosch process to define high-aspect-ratio silicon pillar templates with inherently tapered tips. For the top-tapered configuration, the structures could remain attached to the substrate during the etching process and subsequently be conformally coated using atomic layer deposition (ALD) to form the functional shell. For the bottom-tapered geometry, the patterned nanowire array could be embedded within a membrane and subsequently peeled from the underlying bulk substrate[3]. A controlled sub-micrometer etching step could then be employed to selectively remove the membrane from the designated cylindrical region of the nanowire, thereby exposing the intended active sensing surface. A conformal ALD coating could subsequently be applied to complete the structure.

## Section 2: Signal enhancement for tapered GaP nanowire

### EHN$_{exc}$: GaP tapered NW for different tapering length

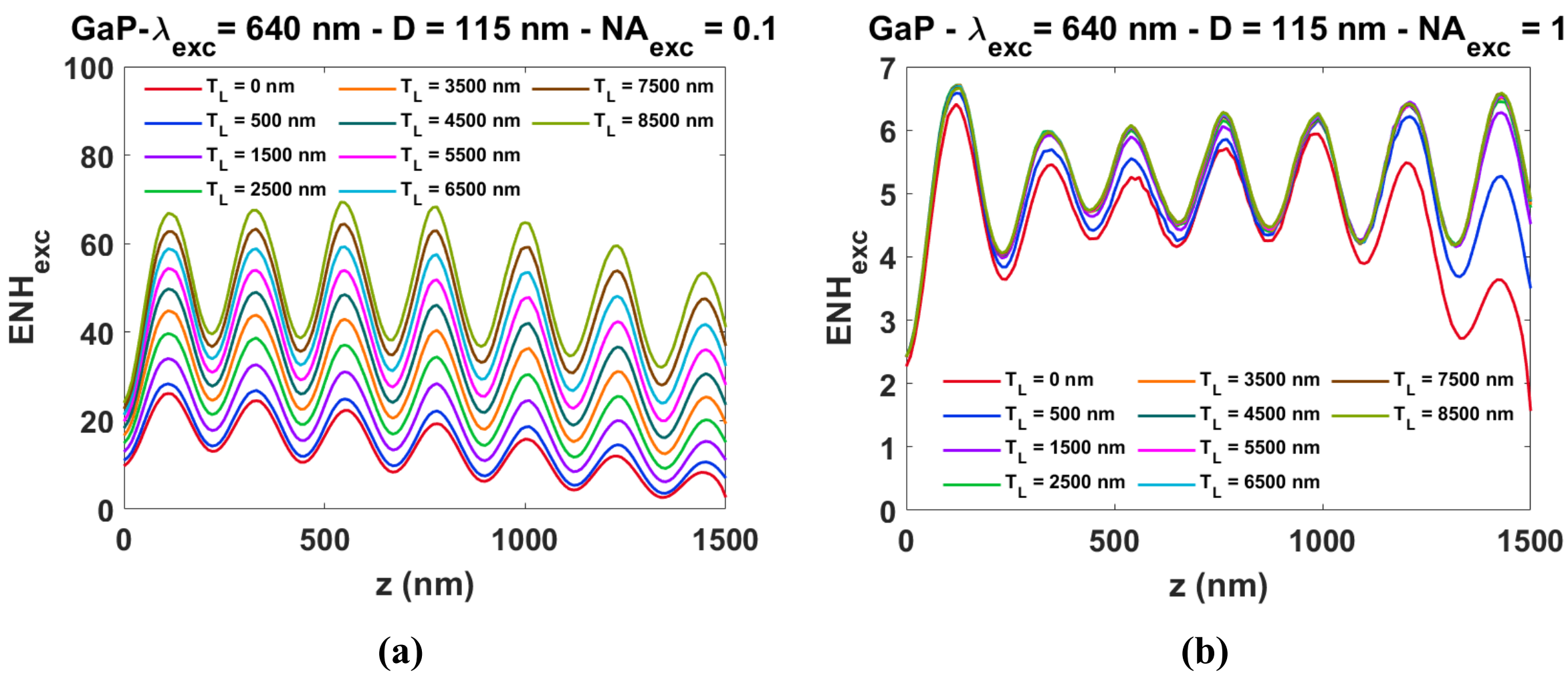

**Figure S22:** EHN$_{exc}$ for tapered GaP nanowire with $C_L$ = 1500 nm, $D$ = 115 nm, and $O_t$ = 10 nm at $\lambda_{exc}$ = 640 nm and $0 \leq T_L \leq 8500$ nm for (a) NA$_{exc}$ = 0.1 and (b) NA$_{exc}$ = 1.

## ⟨$ENH_{exc}$⟩ of tapered GaP nanowire for different tapering length

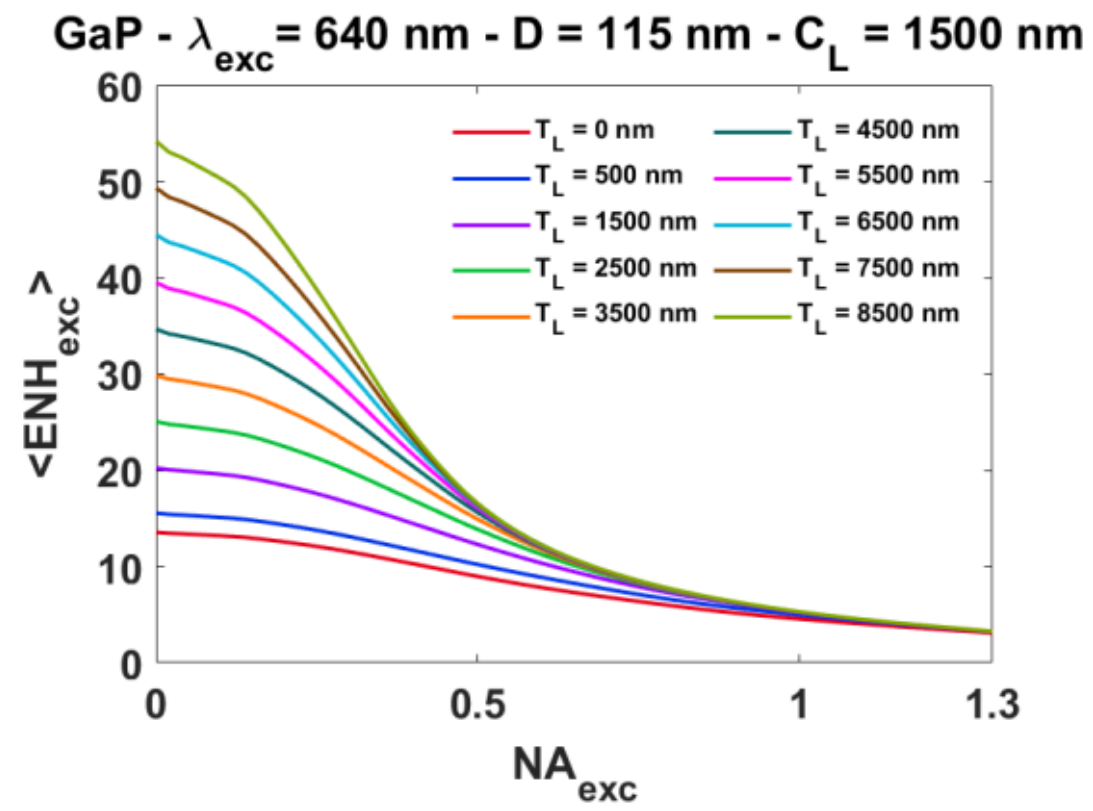

**Figure S23:** ⟨$ENH_{exc}$⟩ for top-tapered GaP nanowire with $C_L$ = 1500 nm, $D$ = 115 nm, and $O_t$ = 10 nm, as a function of $NA_{exc}$ and $0 \leq T_L \leq 8500$ nm.

## ⟨$ENH_{coll}$⟩ of tapered GaP nanowire for different tapering length

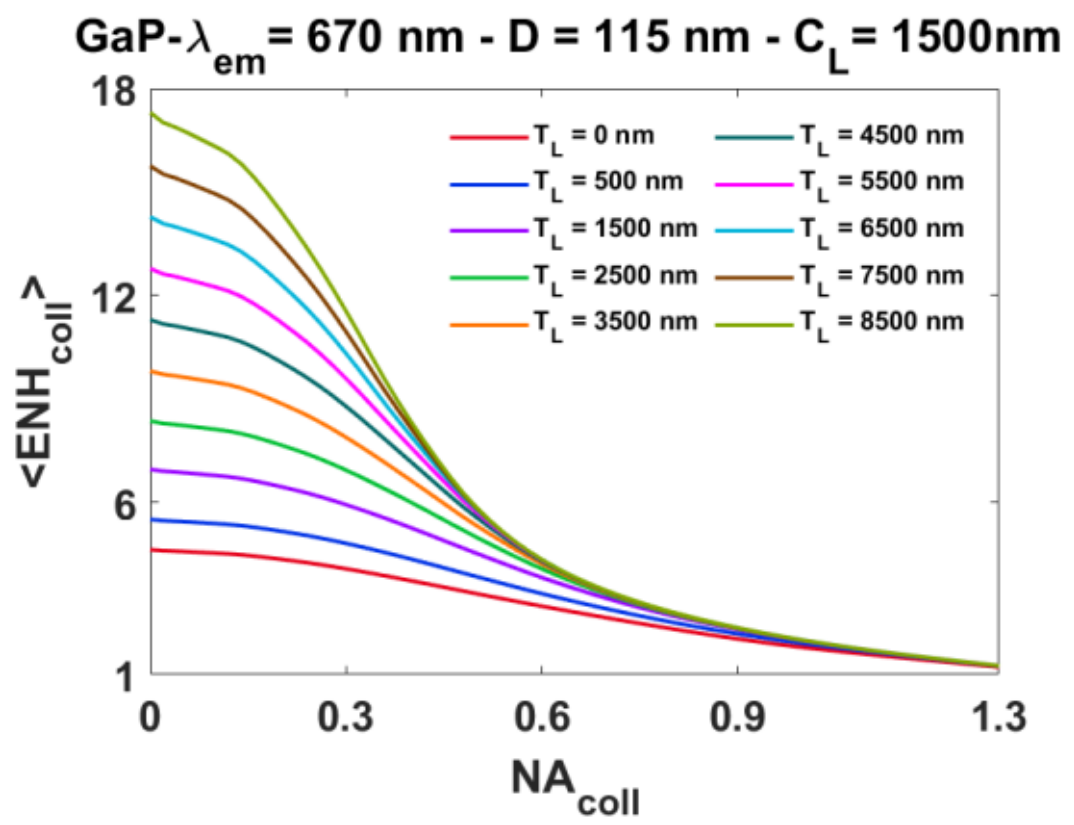

**Figure S24:** ⟨$ENH_{coll}$⟩ for top-tapered GaP nanowires with $C_L$ = 1500 nm, $D$ = 115 nm, $O_t$ = 10 nm, as a function of $NA_{coll}$ for $0 \leq T_L \leq 8500$ nm.

## Purcell Factor:

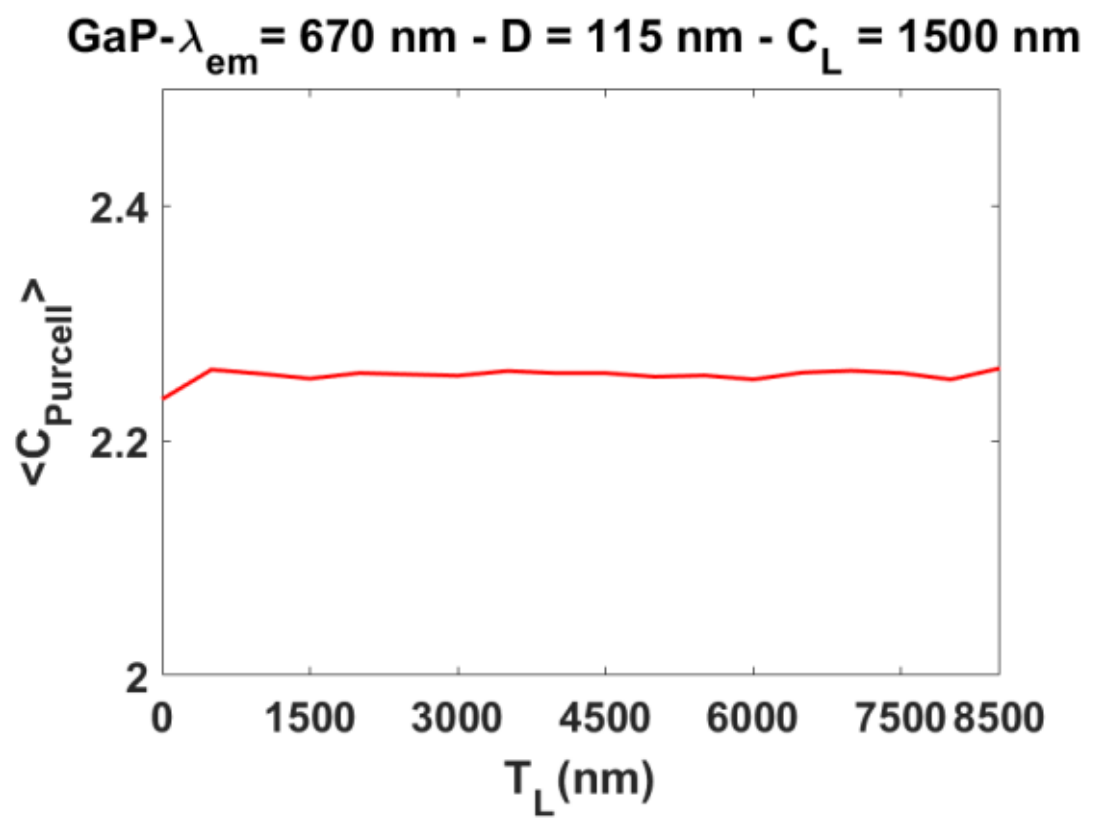

**Figure S25:** Purcell factor of top-tapered GaP nanowire with respect to tapering length

## Signal Enhancement for GaP tapered nanowire:

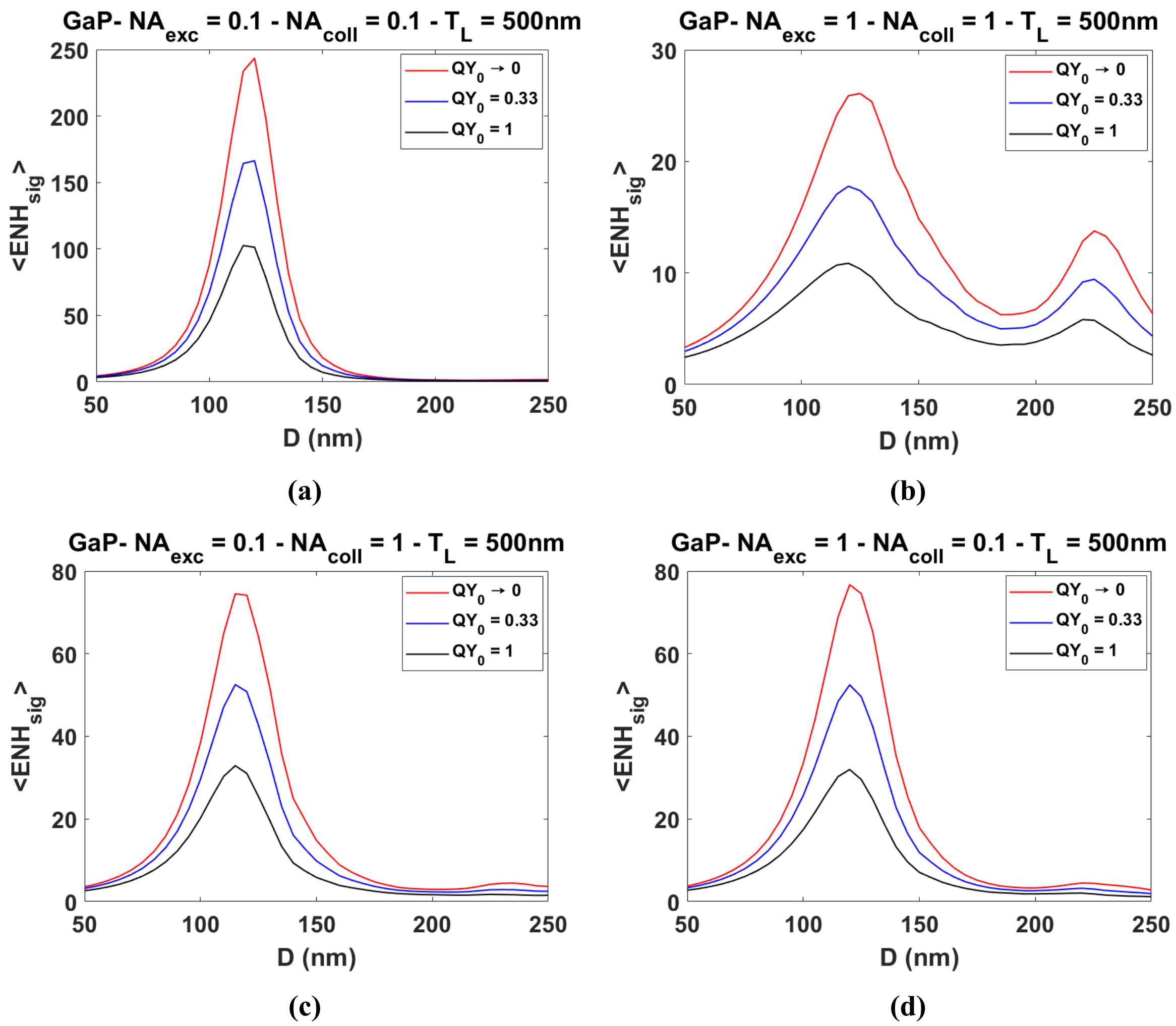

**Figure S26:** ⟨ENHsig⟩ of top-tapered GaP nanowires as a function of nanowire diameter for $QY_0 \rightarrow 0$ , 0.33, and 1 at (a) $NA_{exc} = 0.1$ and $NA_{coll} = 0.1$, (b) $NA_{exc} = 1$ and $NA_{coll} = 1$, (c) $NA_{exc} = 0.1$ and $NA_{coll} = 1$, and (d) $NA_{exc} = 1$ and $NA_{coll} = 0.1$.

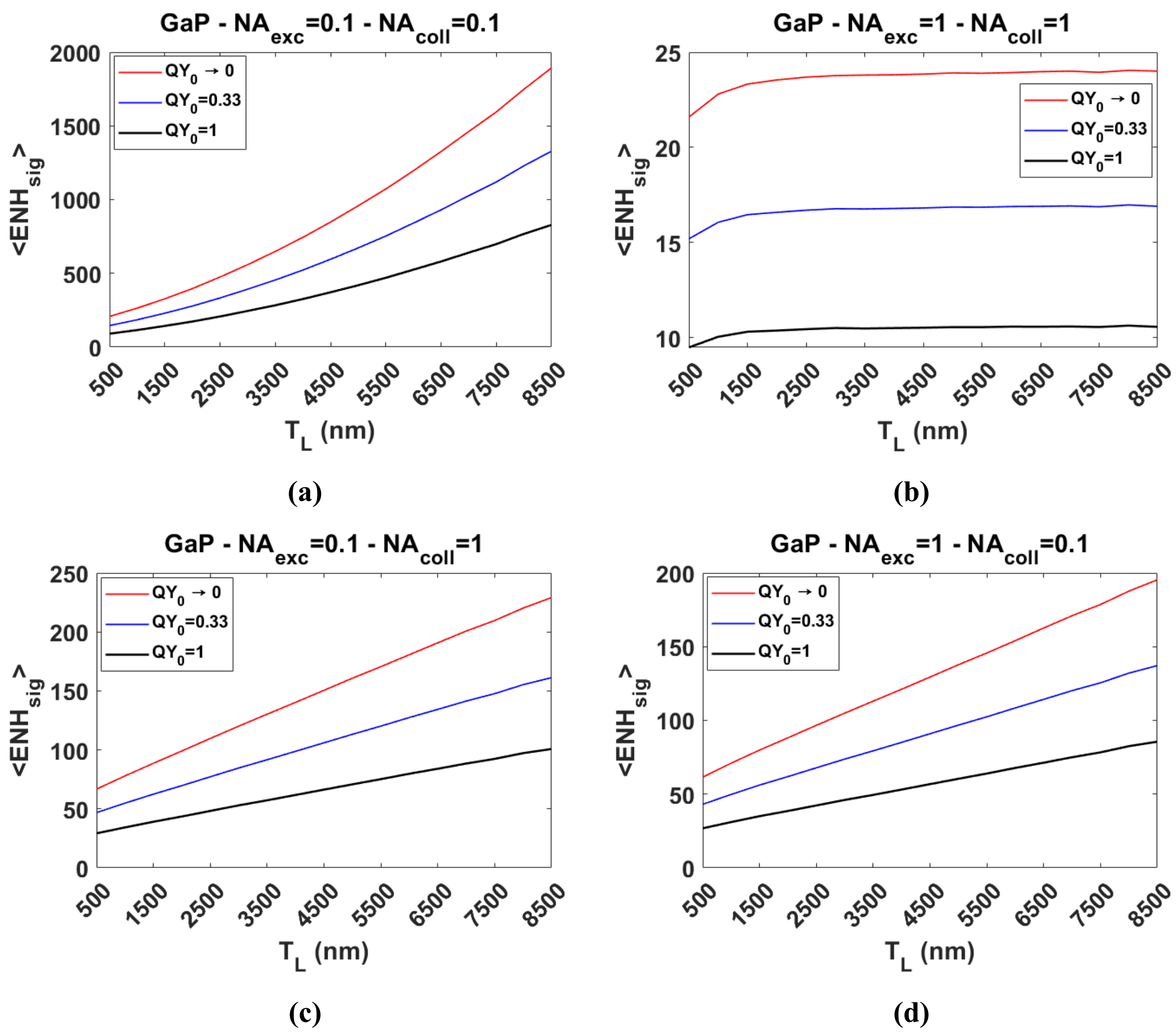

**Figure S27:** ⟨$ENH_{sig}$⟩ of top-tapered GaP nanowires as a function of tapering length ($T_L$) for $QY_0 \rightarrow 0$, 0.33, and 1 at (a) $NA_{exc}$ = 0.1 and $NA_{coll}$ = 0.1, (b) $NA_{exc}$ = 1 and $NA_{coll}$ = 1, (c) $NA_{exc}$ = 0.1 and $NA_{coll}$ = 1, and (d) $NA_{exc}$ = 1 and $NA_{coll}$ = 0.1.